\documentclass[10pt]{article}
\usepackage{graphicx}
\ifx\macrosloaded\relax\endinput\else\let\macrosloaded\relax\fi
\newtheorem{theorem}{Theorem}[section]

\catcode`\@=11
\newcommand{\eqinsec}{\relax\@addtoreset{equation}{section}}
\renewcommand{\theequation}{\ifx\showlabels\iftrue\the\id\else\thesection.\arabic{equation}\fi}
\catcode`\@=13
\newcounter{supeq}
\newenvironment{subeq}
\stepcounter{equation}
\def\theequation{\ifx\showlabels\iftrue\the\id\else\thesection.\arabic{equation}\fi}
\newtoks\id
\newcommand{\eqlabel}[1]{\label{#1}\global\id={(#1)}} 

\newcommand{\tr}{\mbox{tr}}
\newcommand{\be}{\begin{equation}}
\newcommand{\eeq}{\end{equation}}
\newcommand{\bea}{\begin{eqnarray}}
\newcommand{\eea}{\end{eqnarray}}
\newcommand{\beaa}{\begin{eqnarray*}}
\newcommand{\eeaa}{\end{eqnarray*}}
\newcommand{\bseq}{\begin{subeq}}
\newcommand{\eseq}{\end{subeq}}
\newcommand{\ba}{\begin{array}}
\newcommand{\ea}{\end{array}}
\newcommand{\eql}{\eqlabel}

\def \rectangle#1#2{\hbox{\vrule\vbox to #2
{\hrule\hbox to #1{\hfil}\vfil\hrule}\vrule}}
\newcommand{\edd}{\end{document}}

\renewcommand{\c}{\cdot}
\newcommand{\NI}{\noindent}

\newcommand{\Si}{\Sigma}
\newcommand{\ga}{\gamma}

\newcommand{\ggg}{\mbox{${\bf g}$}}

\newcommand{\dd}{\mbox{${\bf D}$}}

\newcommand{\nabb}{\mbox{$\nabla \mkern-13mu /$\,}}

\newcommand{\ddb}{\mbox{$\dd \mkern-13mu /$\,}}

\newtheorem{Le}{Lemma}[section]

\newcommand{\nn}{\nonumber}

\newcommand{\chib}{\underline{\chi}}

\newcommand{\de}{\delta}

\newcommand{\ep}{\epsilon}

\newcommand{\chih}{\hat{\chi}}

\newcommand{\chibh}{\underline{\hat{\chi}}}

\newcommand{\und}[1]{\underline{#1}}

\newcommand{\Nb}{\und{N}}

\newcommand{\Cb}{\und{C}}

\newcommand{\ub}{{\und{u}}}
\renewcommand{\c}{\cdot}

\newcommand{\M}{{\cal M}}

\newcommand{\bb}{\underline{\beta}}
\renewcommand{\a}{\alpha}
\renewcommand{\b}{\beta}

\newcommand{\si}{\sigma}

\newcommand{\ro}{\rho}

\newcommand{\ze}{\zeta}
\newcommand{\divv}{\mbox{div}\mkern-19mu /\,\,\,\,}

\newcommand{\om}{\omega}
\newcommand{\oom}{\Omega}
\newcommand{\oomb}{\underline{\Omega}}
\newcommand{\omb}{\underline{\omega}}

\newcommand{\la}{\lambda}

\newcommand{\dddd}{{\bf D} \mkern-13mu /\,}

\def\frac#1#2{{{#1}\over{#2}}}

\eqinsec
\let\showlabels\iffalse 
\begin{document}
\title{Global characteristic problem for the Einstein vacuum equations with small initial data:
 (II)\ \ \ \ \ \ \ \  The existence proof.}
\author{Giulio Caciotta\ \ \ \  Francesco Nicol\`{o}
\footnote{Dipartimento di Matematica, Universit\`{a} degli Studi di Roma
``Tor Vergata", Via della Ricerca Scientifica, 00133-Roma, Italy}\ \footnote{Part of this research was done at the Isaac Newton
Institute for Mathematical Sciences (Cambridge University), during the semester program ``Global Problems in Mathematical Relativity"
organized by  P.Chrusciel, H.Friedrich, P.Tod.}
\\Universit\`{a} degli studi di Roma ``Tor Vergata"\date{\today}}\maketitle

\begin {abstract} 
{This is the second part of our result on a class of global characteristic problems for the Einstein vacuum equations with small initial data. In
the previous work, \cite{Ca-Ni:char}, denoted by (I), our attention was focused on prescribing the initial data satisfying the costraints imposed by
the characteristic problem. Here we show how the global existence result can be achieved. This part is heavily based on the global results of
D.Christodoulou, S.Klainerman, \cite{C-K:book}, and S.Klainerman, F.Nicol\`o, \cite{Kl-Ni:book}.}
\end{abstract}

\newpage
\section{Introduction}\label{S1}
This is the second paper on the solution of a global characteristic problem for the Einstein vacuum equations with small data. In the first one,
hereafter denoted with (I), we examined the way of posing the initial data on the initial null hypersurfaces, their constraints and their connection
with the initial data given in the harmonic background. Here we show how we can prove an existence result using a strategy which mimics the one used
to prove the global stability of the Minkowski spacetime, \cite{C-K:book}, \cite {Kl-Ni:book}.

\section{The global existence theorem, strategy of the proof.}\label{S2} 
In this section we state the global existence theorem and describe the main steps of its proof. The following
sections are devoted to their implementations. 
\subsection{Statement of the characteristic global existence theorem.}\label{SS2.1}
\begin{theorem}[characteristic global existence theorem]\label{T2.1}
 Let the initial hypersurface be the union of two null truncated cones $C_0\cup\Cb_0$, see figure $1$,
let the initial data \footnote{The null
``cones" and the various quantities which describe the initial data have been defined in(I), section 1.3.} 
\bea
\left\{C_0\ ;\overline{\ga}_{ab},\overline{\oom},\overline{\ze}_a,\overline{\chib}_{ab},\overline{\omb}\right\}\cup\left\{\Cb_0\
;\overline{\ga}_{ab},\overline{\oomb},\overline{\underline X}^a\overline{\ze}_a,\overline{\chi}_{ab},\overline{\om}\right\}\eql{2.54}
\eea 
be assigned together with their partial tangential derivatives to a fixed order specified by the integer $q\geq 7$.
Let us assume they satisfy the smallness conditions:
\bea
J^{(q)}_{C_0\cup\Cb_0}=J^{(q)}_{C_0}\left[\overline{\ga}_{ab},\overline{\oom},\overline{\ze}^a,\overline{\chib},\overline{\omb}\right]
+J^{(q)}_{\Cb_0}\left[\overline{\ga}_{ab},\overline{\oomb},\overline{\ze}_a,\overline{\chi},\overline{\om}\right]\leq
\varepsilon\ ,\eql{2.2w}
\eea
where $J^{(q)}_{C_0}$ and $J^{(q)}_{\Cb_0}$ are defined in {\em(I)}, {\em (I;3.17), (I;3.18)}. Assume, moreover, that the initial data are such that
on $S_0=C_0\cap\Cb_0$, the following estimates for the Riemann null components hold,\footnote{In (I) it is shown that this condition follows 
immediately once some bounds for $\ga,\chi,\chib,\ze$ on $S_0$ are satisfied, see (I), Proposition 2.11 equations (2.91).}
\bea
&&|r_0^{(\frac{7}{2}+\de)-\frac{2}{p}}\b|_{p,S_0}\leq c\varepsilon\ ,\
|r_0^{(3+\de)-\frac{2}{p}}(\ro-\overline{\ro},\si)|_{p,S_0}\leq c\varepsilon\nn\\
&&|r_0^{3-\frac{2}{p}}\ro|_{p,S_0}\leq c\varepsilon\ ,\ |r_0^{(2+\de)-\frac{2}{p}}\bb|_{p,S_0}\leq c\varepsilon\ ,\ \ \ \ \eql{2.3rt}
\eea
where $r_0^2=(4\pi)^{-1}|S_0|$. Then there exists a unique vacuum Einstein spacetime,
$(\M,\ggg)$, solving the characteristic initial value problem with initial data \ref{2.54}. $\M$ is the maximal future Cauchy development of
$C(\la_1)\cup\Cb(\nu_0)$,\footnote{Sometimes to avoid any confusion we write explicitely $C(\la_1;[\nu_0,\infty))\cup\Cb(\nu_0;[\la_1,\la_0])$.
$\M(\nu_1)$ is the maximal Cauchy development of  $C(\la_1;[\nu_0,\nu_1])\cup\Cb(\nu_0;[\la_1,\la_0])$. $C(\la_1)$ and $\Cb(\nu_0)$, the images in
$\M$ of $C_0$ and $\Cb_0$ respectively, are explicitely defined in b). }
\bea
\M=\lim_{\nu_1\rightarrow\infty}\M(\nu_1)=\lim_{\nu_1\rightarrow\infty}J^{+}(S(\la_1,\nu_0))\cap J^{-}(S(\la_0,\nu_1))\ .
\eea
Moreover $\M$ is endowed with the following structures:

 a) $\M$ is foliated by a ``double null canonical foliation" $\{C(\la)\}$, $\{\Cb(\nu)\}$, with $\la\in [\la_1,\la_0]\
,\nu\in[\nu_0,\infty)$.\footnote{recall that with ``double canonical foliation" we refer to a foliation of the spacetime
$(\M,\ggg)$, while with ``canonical foliation" we denote a specific foliation of the initial data on $C_0$. Of course the first one is related to
the second.} Double canonical foliation means that the null hypersurfaces $C(\la)$ are the level hypersurfaces of a function
$u(p)$ solution of the eikonal equation
\[g^{\mu\nu}\partial_{\mu}w\partial_{\nu}w=0\ ,\]
with initial data  a specific function $u_*(p)$ defining the foliation of the ``final" incoming cone,\footnote{The function $u_*(p)$ is the limit as
$\nu_1\rightarrow\infty$ of a function $u_*(\nu_1; p)$ defined on $\Cb(\nu_1;[\la_1,\la_0])$. A detailed discussion of this limit is in
\cite{Kl-Ni:book}, Chapter 8.} while the the null hypersurfaces $\Cb(\nu)$ are the level hypersurfaces of a function
$\ub(p)$ solution of the eikonal equation with initial data a specific function $\ub_{(0)}(p)$ defining the foliation of the initial outgoing cone
$C(\la_1)$.\footnote{We require that $\ub_{(0)}(p)$ defines on $C(\la_1)$ a canonical foliation. In the characteristic case this is
not really needed. This is due to the fact that, as we will discuss later on, the regularity of the initial data is higher than that of the same
quantities in $\M$. We can, nevertheless, require it and, in this case, it is important to remark that the ``canonical" foliation of $C(\la_1)$ is
not the one given when the initial data are specified, see (I). It can be proved that, given the initial data foliation of $C(\la_1)$, it is
possible to build on $C(\la_1)$ a ``canonical" one. Its precise definition and the way for doing it is discussed in \cite{Kl-Ni:book}, Chapter 3 and
with more details in \cite{Niclast}.}
The family of two dimensional surfaces $\{S(\la,\nu)\}$, where
$S(\la,\nu)\equiv C(\la)\cap\Cb(\nu)$, defines a two dimensional foliation of $\M$. 
\smallskip

 b) $i(C_0)=C(\la_1)\ ,\ i(\Cb_0)=\Cb(\nu_0)$, $i(S_0(\nu_0))=i(\underline{S}_0(\la_1))=C(\la_1)\cap\Cb(\nu_0)$\ .

 c) On $C(\la_1)$ with respect to the initial data foliation,\footnote{Observe that while $\cal M$ is globally foliated by a double null canonical
foliation, it is not possible to foliate it with a double foliation solution of the eikonal equation with as initial data the $\oom$-foliation of $C_0$
and $\Cb_0$. This is, nevertheless possible in a small neighbourhood of $C_0$ and $\Cb_0$.} we have
\bea
&&i^*(\ga')=\overline{\ga}\ ,\ i^*(\oom')=\overline{\oom}\ ,\ {i_*}^{-1}(X')=\overline{X}\nn\\
&&i^*(\chi')=\overline{\chi}\ ,\ i^*(\om')=\overline{\om}\ ,\ i^*(\ze')=\overline{\ze}
\eea
together with their tangential derivatives up to $q$. 
\smallskip

 d) On $\Cb(\nu_0)$ with respect to the initial data foliation, we have
\bea
&&i^*(\ga')=\overline{\ga}\ ,\ i^*(\oom')=\overline{\oomb},\ {i_*}^{-1}({\underline X}')=\overline{\underline X}\nn\\
&&i^*(\chib')=\overline{\chib}\ ,\ i^*(\omb')=\overline{\omb}\ ,\ i^*(\ze')=\overline{\ze}
\eea
where $\ga',\oom',X',\chi',\chib',\om',\omb',\ze'$ are the metric components and the connection coefficients in a neighbourhood
of $C(\la_1)$ and $\Cb(\nu_0)$.\footnote{These quantities denoted here $\ga',\oom',......$ were denoted in (I)
$\ga,\oom,\oomb,X,{\underline X}$, $\chi,\chib,\om,\omb,\ze$. In the sequel these notations will be referred to the corresponding quantities relative
to the double canonical foliation of $\M$ and their restrictions to $C_0\cup\Cb_0$.}
\smallskip
 
 e) The costraint equations {\em (I;1.55),(I;1.56)} are the pull back of (some of) the structure equations of $\M$
restricted to $C(\la_1)$ and $\Cb(\nu_0)$.\footnote{With respect to the foliations of the initial data.}
\smallskip

 f) Let the double null canonical foliation and the associated two dimensional one, $\{S(\la,\nu)\}$ be given. We can define on the whole $\M$ a
null orthonormal frame $\{e_4,e_3,e_a\}$ adapted to it, the metric components, $\ga_{ab},\oom,X^a$ with respect to
the adapted coordinates $\{u,\ub,\theta,\phi\}$\footnote{The precise way the coordinate $\theta$ and $\phi$ are defined will be discussed elsewhere.
See also \cite{Kl-Ni:book}, paragraph 3.1.6.} and the corresponding connection coefficients
\bea
\chi_{ab}&=&\ggg(\dd_{e_a}e_4,e_b)\ \ ,\ \chib_{ab}=\ggg(\dd_{e_a}e_3,e_b)\nn\\
\omb&=&\frac{1}{4}\ggg(\dd_{e_3}e_3,e_4)=-\frac{1}{2}\dd_3(\log\oom)\nn\\
\om&=&\frac{1}{4}\ggg(\dd_{e_4}e_4,e_3)=-\frac{1}{2}\dd_4(\log\oom)\eql{4.16ww}\\
\ze_{a}&=&\frac{1}{2}\ggg(\dd_{e_{a}}e_4,e_3)\ .\nn
\eea
Moreover $X^a$ is a vector field defined in $\M$ such that denoting $N$ and $\Nb$ two null vector fields equivariant with respects to
the $S(\la,\nu)$ surfaces, the following relations hold: $N=\oom e_4$, $\Nb=\oom e_3$ and in the $\{u,\ub,\theta,\phi\}$ coordinates 
\[N=2\oom^2\frac{\partial}{\partial\ub}\ ,\ \Nb=2\oom^2\left(\frac{\partial}{\partial u}+X^a\frac{\partial}{\partial\om^a}\right)\ .\]

 g) The null geodesics of $\M$ along the outgoing and incoming null direction
$e_4,e_3$ are defined for all $\nu\in[\nu_0,\infty)$ and all $\la\in[\la_1,\la_0]$ respectively.
Finally the existence result is uniform in $\la_1(<0)$.\footnote{This means that the quantities \ref{2.2w}, depending on the initial data, which
we require to be small are independent from $\la_1$.} 
\end{theorem}

\subsection{The general structure of the proof.}\label{SS2.2}
The proof is made by two different parts whose structure is, basically, the same.
\smallskip

{\bf First part:} We prove the existence of the spacetime
\[\M'=J^+(S(\la_1,\nu_0))\cap J^-(S(\la_0,\nu_1))\ ,\]
{where}, see figure 2,\footnote{All the null hypersurfaces are represented as if the spacetime $\M$ were flat.} $\ \ \ \ \ \ \
2\ro_0=\nu_0-\la_1=\nu_1-\la_0\ .$ 

\medskip

{\bf Second part:} We prove the existence of the spacetime $\M\supset\M',$
\[\M=\lim_{\nu_1\rightarrow\infty}J^+(S(\la_1,\nu_0))\cap J^-(S(\la_0,\nu_1))\ ,\]
see figure 3.

{\bf Remark:} {It is appropriate at this point to present an intuitive picture of the spacetime we are building. As our result is a small data result we
expect that our spacetime stays near to (a portion of) the Minkowski spacetime. This implies that the functions $\ub(p)$ and $u(p)$ do not differ
much, written in spherical Minkowski coordinates, from $t+r$ and $t-r$, respectively. Moreover the radius $r(\la,\nu)$ defined as
$(4\pi)^{-\frac{1}{2}}$ times the square root of the area of $S(\la,\nu)$, see {\em (I;2.13)}, will stay near the $r$ spherical coordinate. The
initial cones $C_0$ and $\Cb_0$ approximate two truncated minkowskian cones, one outgoing and one incoming, with their vertices on the origin
vertical axis. The assumption that $\nu_0$ and $|\la_1|$ are approximately equal to $r_0$ just pictures the two dimensional surface $S_0$ as lying
(approximately) on the hyperplane $t=0$ where $t$ is ``near" to $\frac{1}{2}(\ub+u)$.}\footnote{Although the picture defined here is the more
natural, with a redefinition of the functions $\ub,u,r$ in terms of the standard coordinates one could also treat in a similar way the case where the
two ``initial" cones $C_0$ and $\Cb_0$ are shifted one respect to the other and their vertices do not lie even approximatley on the same vertical
axis.}
\medskip

 {\bf Proof of the first part:}
We denote $({\cal K}(\tau),\ggg)$ a solution of the ``characteristic Cauchy problem" with the following properties:
\smallskip

{\bf i)} $({\cal K}(\tau),\ggg)$ is foliated by a double null canonical foliation $\{C(\la)\}$, $\{\Cb(\nu)\}$
with $\la\in [\la_1,{\overline\la}]\ ,\ \nu\in [\nu_0,{\overline\nu}]$. Moreover
\[i(C_0)\cap{\cal K}(\tau)=C(\la_1;[\nu_0,{\overline\nu}])\ ;\  i(\Cb_0)\cap{\cal K}(\tau)=\Cb(\nu_0;[\la_1,{\overline\la}])\]
where \footnote{As the spacetime we obtain is ``near" to a portion of the Minkowski spacetime,
condition \ref{3.22ggw} can be visualized as the requirement that $S({\overline\la},{\overline\nu})$ coincides with a vertical (time) translation of
$S_0$.} 
\bea
{\overline\nu}+{\overline\la}=2\tau\ \ \ \ \ ;\ \ \ \ \ {\overline\nu}-{\overline\la}=2\ro_0\ . \eql{3.22ggw}
\eea

{\bf ii)}
Denoted $\{e_3,e_4,e_a\}$ the null orthonormal frame adapted to the double null canonical foliation. We introduce a family of norms ${\cal R} , {\cal
O}$ for the null components of the Riemann curvature tensor and for the connection coefficients respectively, as done in \cite{Kl-Ni:book},
Chapter 3. 
\smallskip

{\bf iii)} Given $\ep_0>0$ sufficiently small, but larger than $\varepsilon$, the norms ${\cal R} , {\cal O}$ satisfy the
following inequalities
\bea
{\cal R}\leq \ep_0\ \ ,\ \ {\cal O}\leq \ep_0\ .
\eea

{\bf iv)} Denoted by $\cal T$ the set of all values $\tau$ for which the spacetime
${\cal K}(\tau)$ does exist, we define $\tau_*$ as the $\sup$ over all the values of $\tau\in{\cal T}$, see figure 4:
\bea
\tau_* =\sup \{\tau\in {\cal T}\}\eql{2.58}\ .
\eea
There are, now, two possibilities:
\bea
\tau_*=\la_0+\ro_0 \ \ \ \ \mbox{or}\ \ \ \ \tau_*<\la_0+\ro_0\ \ .\eql{2.10w}
\eea
If $\tau_*=\la_0+\ro_0$ then 
\bea
{\cal K}(\tau_*)=J^+(S(\la_1,\nu_0))\cap J^-(S(\la_0,\nu_1))
\eea 
and the first part is proved. If, viceversa, $\tau_*<\la_0+\ro_0$,
we show that in ${\cal K}(\tau_*)$ the norms $\cal R$ and $\cal O$ satisfy the following estimates
\bea
{\cal R}\leq \frac{1}{2}\ep_0\ \ ;\ \ {\cal O}\leq \frac{1}{2}\ep_0\ .\eql{2.59}
\eea

{\bf v)} Using the inequalities \ref{2.59} restricted to $\Cb(\nu_*)$ it is possible to prove that we can extend the
spacetime
${\cal K}(\tau_*)$ to a spacetime ${\cal K}(\tau_*+\de)$. This contradicts the fact that $\tau_*$ is the supremum defined in
\ref{2.58}, unless $\tau_*=\la_0+\ro_0$. In fact in this case ${\cal K}(\tau_*)$ describes the maximal region ${\cal
K}(\tau)$ which can be determined from the initial conditions assigned on $\Cb(\nu_0;[\la_1,\la_0])\cup
C(\la_1;[\nu_0,\nu_1])$ even if ${\cal R}\leq \frac{1}{2}\ep_0 ;\ {\cal O}\leq \frac{1}{2}\ep_0$.



\smallskip

{\bf vi)} Due to the fact, which will be evident in the course of the proof, that the estimates for
$\cal R$ and $\cal O$ do not depend on the magnitude of the interval $[\la_1,\la_0]$, also the proof of the first part can be
considered a global existence result. In other words fixed $\la_0$, we can choose $\ro_0=\frac{1}{2}(\nu_0-\la_1)$
arbitrarily large which implies that also $\tau_*=\la_0+\ro_0$ can be chosen arbitrarily large.
\smallskip

\NI In conclusion the result of the first part is achieved if we prove that:
\smallskip

i) The set ${\cal T}$ is not empty.
\smallskip

ii) In ${\cal K}(\tau_*)$ the norms $\cal R$ and $\cal O$ satisfy the following estimates
\bea
{\cal R}\leq \frac{1}{2}\ep_0\ \ ;\ \ {\cal O}\leq \frac{1}{2}\ep_0\ .\eql{2.59a}
\eea

iii) If $\tau_*<\la_0+\ro_0$ we can extend ${\cal K}(\tau_*)$ to ${\cal K}(\tau_*+\de)$ with $\de>0$\ .
\bigskip

\NI Steps i), ii) and iii) are the technical parts of the result. Their detailed proofs are in the subsequent sections.
\smallskip

\NI {\bf Proof of the second part:} It is, basically, a corollary of the proof of the first part.
We consider in this case the spacetimes $\M({\overline\nu})$ solutions of the ``characteristic Cauchy problem" (or the
``characteristic canonical Cauchy problem") with the following properties:
\smallskip

{\bf i)} $({\M}(\nu),\ggg)$ is foliated by a double null canonical foliation $\{C(\la)\}$, $\{\Cb(\nu)\}$
with
$\la\in [\la_1,\la_0]\ ,\ \nu\in [\nu_0,{\overline\nu}]$. Moreover
\[i(C_0)\cap{\M}({\overline\nu})=C(\la;[\nu_0,{\overline\nu}])\ ;\ 
i(\Cb_0)\cap{\M}({\overline\nu})=\Cb(\nu_0;[\la_1,\la_0])\] 
\smallskip 

{\bf ii)}
let $\{e_3,e_4,e_a\}$ be the null orthonormal frame adapted to the double null canonical foliation, defined as in
\cite{Kl-Ni:book}, Chapter 3. We introduce a family of norms ${\cal R} , {\cal O}$ for the connection coefficients and for
the null components of the Riemann curvature tensor respectively, as done in \cite{Kl-Ni:book}, Chapter 3. 
\smallskip

{\bf iii)} Given $\ep_0>0$ sufficiently small, but larger than $\varepsilon$, the norms ${\cal R} , {\cal O}$ satisfy the
following inequalities
\bea
{\cal R}\leq \ep_0\ \ ,\ \ {\cal O}\leq \ep_0\ .
\eea

{\bf iv)} Let $\cal N$ be the set of all values $\overline\nu$ for which the spacetime
$\M(\overline\nu)$ does exist. We define $\nu_*$ as the $\sup$ over all the values of $\overline\nu\in{\cal N}$, see figure 5:
\bea
\nu_* =\sup\{\overline\nu\in {\cal N}\}\eql{2.58a}\ .
\eea
Again there are two possibilities: if $\nu_*=\infty$ the result is achieved, therefore we are left to show that the
second possibility, $\nu_*<\infty$, leads to a contradiction. For it we show first that in $\M(\nu_*)$ the norms $\cal R$
and $\cal O$ satisfy the following estimates
\bea
{\cal R}\leq \frac{1}{2}\ep_0\ \ ;\ \ {\cal O}\leq \frac{1}{2}\ep_0\ .\eql{2.59aa}
\eea

{\bf v)} Using inequalities \ref{2.59aa} restricted to $\Cb(\nu_*;[\la_1,\la_0])$, it is possible to prove that we can
extend the spacetime $\M(\nu_*)$ to a spacetime ${\M}(\nu_*+\de)$, a fact which contradicts the definition of $\nu_*$ unless
$\nu_*=\infty$\ .
\smallskip

\NI{\bf Remark:} Observe that in the second part of the proof we do not have to show that the set $\cal N$ is not empty as, due
to the first part of the proof,  $\cal N$ contains, at least, the element $\overline\nu=\la_0+2\ro_0$\ .

\subsection{A broad sketch of the technical parts of the proof.}\label{SS2.3}

{\bf First part:}
\smallskip

{\bf i)} To prove that the set $\cal T$ is not empty we have to show that a spacetime ${\cal K}(\tau)$ exists, possibly with
small $\tau$. This requires a local existence theorem and we can borrow it from the results of A.Rendall,
\cite{rendall:charact} or of H.Muller Zum Hagen, H.Muller Zum Hagen and H.J.Seifert, \cite{Muller}, \cite{MullerSeifert}. The main difficulty in adapting
these results to the present case is that they are proved using harmonic coordinates while the ``gauge" we use for ${\cal K}(\tau)$ is
the one associated to the double null canonical foliation.
This requires a precise connection between the initial data written in the two different gauges in such a way
that we can reexpress their results in our formalism. This will be discussed in a next section.
\smallskip

{\bf ii)} Once we have proved that $\cal T$ is not empty we can define ${\cal K}(\tau_*)$ and prove that in this spacetime
inequalities \ref{2.59},
\[{\cal R}\leq \frac{1}{2}\ep_0\ \ ;\ \ {\cal O}\leq \frac{1}{2}\ep_0\ ,\]
 hold. This requires, see \cite{Kl-Ni:book}, Chapter 3 for a  detailed discussion,
that the double null foliation of ${\cal K}(\tau_*)$ be canonical which, at its turn, requires to prove the existence of specific foliations,
denoted again ``canonical", on $C(\la_1)$ and on $\Cb(\nu_*)$. These ``canonical" foliations will play the role of the ``initial data" for the
solutions of the eikonal equations
\[g^{\mu\nu}\partial_{\mu}w\partial_{\nu}w=0\]
whose level hypersurfaces define the double null canonical foliation $\{C(\la)\}$, $\{\Cb(\nu)\}$ of ${\cal K}(\tau_*)$, see also
\cite{Niclast}.\footnote{The existence of this foliation has to be proved also for the (local) spacetime of i). The proof one gives for ${\cal
K}(\tau_*)$ can be easily adapted to this case. The canonical foliation is crucial to obtain on $\Cb(\nu_*)$ all the connection coefficients, 
to be used again as initial data for the backward transport equations, without any loss of derivatives.}
\medskip

{\bf iii)} The central part of the proof consists in showing that we can extend ${\cal K}(\tau_*)$ to ${\cal K}(\tau_*+\de)$ with
$\de>0$, see figure 6.

To achieve it we have to implement the following steps:
\smallskip

 1) First of all we have to prove an existence theorem for the strips $\Delta_1{\cal K}\ ,\ \Delta_2{\cal K}$,
\bea
&&\Delta_1{\cal K}=J^+(S(\la_1,\nu_*)\cap J^-(S({\overline\la}(\nu_*),\nu_*+\de))\eql{3.31wq}\\
&&\Delta_2{\cal K}=J^+(S({\overline\la}(\nu_*),\nu_0)\cap J^-(S({\overline\la}(\nu_*+\de),\nu_*))\ ,\nn
\eea
where, see \ref{2.10w},\footnote{From \ref{2.10w} and \ref{2.19y} it follows that $\la_0-\overline{\la}(\nu_*)\geq 0$.}
 \bea{\overline\la}(\nu)=-2\ro_0+\nu\ .\eql{2.19y}\eea
and whose boundaries are:
\bea
\partial\Delta_1{\cal K}\!&=&\!C(\la_1;[\nu_*,\nu_*+\de])\cup\Cb(\nu_*+\de;[\la_1,{\overline\la}(\nu_*)])\eql{aa}\\
&&\cup\,C({\overline\la}(\nu_*);[\nu_*,\nu_*+\de])\cup\Cb(\nu_*;[\la_1,{\overline\la}(\nu_*)])\nn
\eea
\bea
\partial\Delta_2{\cal K}\!&=&\!\Cb(\nu_0;[{\overline\la}(\nu_*),{\overline\la}(\nu_*+\de)])\cup
C({\overline\la}(\nu_*+\de);[\nu_0,\nu_*])\\
&&\cup\,\Cb(\nu_*;[{\overline\la}(\nu_*),{\overline\la}(\nu_*+\de)])\cup C({\overline\la}(\nu_*);[\nu_0,\nu_*])\ ,\nn
\eea
\smallskip

 2) We are then left with proving the (local) existence of a diamond shaped region,
\bea
\Delta_3{\cal K}=J^{+}(S({\overline\la}(\nu_*),\nu_*)\cap J^{-}(S({\overline\la}(\nu_*+\de),\nu_*+\de))\ ,
\eea
specified by the initial data
\bea
C({\overline\la}(\nu_*);[\nu_*,\nu_*+\de])\cup\Cb(\nu_*;[{\overline\la}(\nu_*),{\overline\la}(\nu_*+\de)])\ .
\eea

 3) Finally on the portion of the boundary of
\[{\cal K}(\tau_*+\de)={\cal K}(\tau_*)\cup\Delta_1{\cal K}\cup\Delta_2{\cal K}\cup\Delta_3{\cal K}\ ,\]
made by $\Cb(\nu_*+\de;[\la_1,{\overline\la}(\nu_*+\de))])$ \footnote{ The canonical foliation on $C(\la_1;[\nu_0,\nu_*+\de])$
was already proved and it does not change.} a new canonical foliation has to be constructed which stays ``near" to the (non canonical) foliation
obtained extending the double null canonical foliation of ${\cal K}(\tau_*)$ up to $\Cb(\nu_*+\de;[\la_1,{\overline\la}(\nu_*+\de))])$ which will be
considered, in this case, as the background foliation.

\NI This completes the description of all the steps needed to prove our result. The detailed proof will be given in the following sections.
\medskip

\NI{\bf Second part:}

In this case, as already said, we know that the set $\cal N$ is not empty. To prove that $\nu_*=\infty$ we have to show
that, if $\nu_*$ were finite, the spacetime
${\M}(\nu_*)$ could be extended to a spacetime ${\M}(\nu_*+\de)$ with the same properties. To prove it we need an
existence theorem for the strip, see figure 5,
\[\Delta\M=J^+(S(\la_1,\nu_*))\cap J^{-}(S(\la_0,\nu_*+\de))\ .\]
As in the case of $\Delta_1{\cal K}$ and $\Delta_2{\cal K}$, this existence theorem is a non local result as the
``length" of the strip does not have to be small in the $e_3$ direction. Therefore also in this case the proof requires a bootstrap
mechanism. 

\NI Finally, as at the end of the proof of the first part, we have to show that $\M(\nu_*+\de)$ can be endowed with a double
canonical foliation whose existence can be proved once we prove the existence of a canonical foliation on the ``last slice",
$\Cb(\nu_*+\de;[\la_1,\la_0])$, see \cite{Kl-Ni:book} Chapter 3 and \cite{Niclast}.
\smallskip

\NI {\bf Remark: }{One has to keep in mind that moving from ${\cal K}(\tau_*)$ to ${\cal K}(\tau_*+\de)$ the double null canonical
foliation of ${\cal K}(\tau_*+\de)$ 
is  different from to the one associated to ${\cal K}(\tau_*)$. The difference is in the outgoing cones $\{C(\la)\}$. In fact the
ingoing cones $\{\Cb(\nu)\}$ of ${\cal K}(\tau_*+\de)$ are just the extension (in $\la$) of those already present in ${\cal K}(\tau_*)$, as 
they are the level surfaces of the function $\ub(p)$ solution of the eikonal equation with initial data on $C(\la_1)$ (associated to its canonical
foliation). The situation is different for the outgoing cones $\{C(\la)\}$. They are also the level hypersurfaces of a solution $u(p)$ of
the eikonal equation, but this time the initial data are associated to the canonical foliation of the upper boundary, $\Cb(\nu_*)$ and
$\Cb(\nu_*+\de)$ respectively. Therefore the outgoing cones of the double null canonical foliation of ${\cal K}(\tau_*+\de)$ are not an extension of
the corresponding ones of ${\cal K}(\tau_*)$. Nevertheless it can be proved easily that they differ from the previous ones by a quantity\footnote{To
be precise we have to define a quantity describing the nearness of the two families of cones, for instance one can define it as the $\sup$ of the
oscillation of the new function $u(p)$ along the previous $S(\la,\nu)$.} which is small and does not depend on the length of the cones,
see for instance \cite{Kl-Ni:book} Chapter 8.}

\section{Technical parts of the existence proof}\label{S.3}
As it will be clear in the course of the proof, the whole program to be completed requires a local existence proof for the characteristic
problem; this is somewhat a separate step which can be done in many ways, therefore we state only the result we need and 
postpone the discussion of it to a subsequent section.
\subsection{Local existence result}\label{SS3.1}
In the proof of our result a bootstrap argument is used many times. Each time to implement it a local existence result is needed.
We show that all these local results are of the same type.

\NI The first time a local existence is needed is when we prove that the set $\cal T$ is not empty. This requires that, possibly for small
${\overline\tau}$, a (portion of) spacetime, ${\cal K}({\overline\tau})$ must exist, satisfying the properties i),...,vi) given at the
beginning of subsection \ref{SS2.2}. This follows from an existence proof for a characteristic Cauchy problem with
initial data on the union of two null hypersurfaces, one corresponding to an ``outgoing cone",
$C(\la_1)$ and the other one to an incoming cone $\Cb(\nu_0)$, intersecting along the two dimensional surface $S_0=S(\la_1,\nu_0)$.

\NI The region ${\cal K}({\overline\tau})$ must be, given ${\overline\tau}=\frac{1}{2}(\overline{\nu}+\overline{\la})$, the maximal Cauchy
development associated to
the initial null hypersurface 
${\cal C}_{\overline{\tau}}=C(\la_1;[\nu_0,\overline{\nu}])\cup\Cb(\nu_0;[\la_1,\overline{\la}])$. Therefore the boundary of ${\cal
K}({\overline\tau})$ will be:
\bea
\partial{\cal K}({\overline\tau})=C(\la_1;[\nu_0,\overline{\nu}])\cup\Cb(\nu_0;[\la_1,\overline{\la}])
\cup\Cb({\overline\nu};[\la_1,\overline{\la}])\cup C(\overline{\la};[\nu_0,\overline{\nu}])
\eea
with $\overline{\la}, \overline{\nu}$ satisfying\ \ \ $\overline{\nu}-\overline{\la}=2\ro_0$\ .
\smallskip

\NI In the local existence proof the initial data norms on ${\cal C}_{\overline\tau}$ must be sufficiently small, $\leq\varepsilon$, so that the
corresponding norms on 
$\Cb({\overline\nu};[\la_1,\overline{\la}])\cup C(\overline{\la};[\nu_0,\overline{\nu}])$ are bounded by $c\varepsilon$ with
$\varepsilon$ such that $c\varepsilon\leq\ep_0$.

\NI Once this result is proved the set $\cal T$ is not empty and we can define ${\cal K}(\tau_*)$. Next step consists in showing that the
region ${\cal K}(\tau_*)$ can be extended to a larger region ${\cal K}(\tau_*+\de)$ endowed with the same properties. As discussed in the
following subsections this requires proving the existence of the slab $\Delta_1{\cal K}\cup\Delta_2{\cal K}\cup\Delta_3{\cal K}$.

\NI Let us consider first the portion $\Delta_1{\cal K}$. As discussed in subsection \ref{SS.3.4}, to prove its existence a bootstrap
argument is required. Again this requires a local existence result with initial data on the null hypersurface ${\cal
C}_1=C(\la_1;[{\overline\nu},{\overline\nu}+\de])\cup\Cb({\overline\nu};[\la_1,\la_1\!-\!{\de}])$, whose norms are bounded by
$c_1\varepsilon$. The same local existence proof is required for the region $\Delta_2{\cal K}$ and finally, if $\de$ is sufficiently small, a
local existence proof is enough to prove the existence of the region $\Delta_3{\cal K}$. 

\NI The conclusion is that the various local existence results we need are always of the same type which can be summarized in the following
theorem:

\begin{theorem}[Local existence result]\label{theor3.1}
Let the initial data 
\bea
\left\{C_{in}\ ;\overline{\ga}_{ab},\overline{\oom},\overline{\ze}_a,\overline{\chib}_{ab},\overline{\omb}\right\}\cup\left\{\Cb_{in}\
;\overline{\ga}_{ab},\overline{\oomb},\overline{\underline X}^a\overline{\ze}_a,\overline{\chi}_{ab},\overline{\om}\right\}\eql{2.54z}
\eea 
be assigned together with their partial tangential derivatives to a fixed order specified by the integer $q\geq 7$.
Let us assume they satisfy the smallness conditions:
\bea
J^{(q)}_{C_{in}\cup\Cb_{in}}=J^{(q)}_{C_{in}}\left[\overline{\ga}_{ab},\overline{\oom},\overline{\ze}^a,\overline{\chib},\overline{\omb}\right]
+J^{(q)}_{\Cb_{in}}\left[\overline{\ga}_{ab},\overline{\oomb},\overline{\ze}_a,\overline{\chi},\overline{\om}\right]\leq
\varepsilon\ ,
\eea
where $J^{(q)}_{C_{in}}$ and $J^{(q)}_{\Cb_{in}}$ are defined as $J^{(q)}_{C_{0}}$ and $J^{(q)}_{\Cb_{0}}$ in {\em(I)}, {\em (I;3.17) , (I;3.18)}.
Then there exists a couple $({\la}_{fin},{\nu}_{fin})$ satisfying
\[{\nu}_{fin}-{\la}_{fin}=2\overline\ro\ \ \ ,\ \ {\nu}_{fin}+{\la}_{fin}=2{\overline\tau}\ \ ,\]
with ${\overline\tau}=\tau_{in}+\overline\si$, with $\overline\si$ a positive number, possibly small, such that a vacuum Einstein spacetime,
$({\cal K}({\overline\tau}),\ggg)$ exists, the maximal future development of
$C(\la_{in};[\nu_{in},{\nu}_{fin}])\cup\Cb(\nu_{in};[\la_{in},{\la}_{fin}])$,
\bea
{\cal K}({\overline\tau})=J^{+}(S(\la_{in},{\nu_{in}}))\cap J^{-}(S({\la}_{fin},\nu_{fin}))\ .
\eea
$C(\la_{in};[\nu_{in},{\nu}_{fin}])\cup\Cb(\nu_{in};[\la_{in},{\la}_{fin}])$ is the image, relative to an immersion $i$, of a
portion of the initial data null hypersurface ${\cal C}_{in}=C_{in}\cup\Cb_{in}$ and the couple $(\la_{in},\nu_{in})$ satisfies:
\[{\nu}_{in}-{\la}_{in}=2\overline\ro\ \ \ ,\ \ {\nu}_{in}+{\la}_{in}=2{\tau_{in}}\ .\] 
Moreover ${\cal K}({\overline\tau})$ can be endowed with the following structures:
\smallskip

 a) A ``double null canonical foliation" $\{C(\la;[\nu_{in},{\nu}_{fin}])\}$, $\{\Cb(\nu;[\la_{in},{\la}_{fin}])\}$, with $\la\in
[\la_{in},{\la}_{fin}]\ ,\nu\in[\nu_{in},{\nu}_{fin}]$. The family of two dimensional surfaces
$\{S(\la,\nu)\}$, where
$S(\la,\nu)\equiv C(\la;[\nu_{in},{\nu}_{fin}])\cap\Cb(\nu;[\la_{in},{\la}_{fin}])$, defines a two dimensional foliation of ${\cal
K}({\overline\tau})$. 
\smallskip

 b) On $C(\la_{in};[\nu_{in},{\nu}_{fin}])$ with respect to the initial data foliation, we have
\bea
&&i^*(\ga')=\overline{\ga}\ ,\ i^*(\oom')=\overline{\oom}\ ,\ {i_*}^{-1}(X')=\overline{X}\nn\\
&&i^*(\chi')=\overline{\chi}\ ,\ i^*(\om')=\overline{\om}\ ,\ i^*(\ze')=\overline{\ze}
\eea
together with their tangential derivatives up to $q$. 
\smallskip

 c) On $\Cb(\nu_{in};[\la_{in},{\la}_{fin}])$ with respect to the initial data foliation, we have
\bea
&&i^*(\ga')=\overline{\ga}\ ,\ i^*(\oom')=\overline{\oomb},\ {i_*}^{-1}({\underline X}')=\overline{\underline X}\nn\\
&&i^*(\chib')=\overline{\chib}\ ,\ i^*(\omb')=\overline{\omb}\ ,\ i^*(\ze')=\overline{\ze}
\eea
where $\ga',\oom',X',\chi',\chib',\om',\omb',\ze'$ are the metric components and the connection coefficients in a neighbourhood
of $C(\la_{in};[\nu_{in},{\nu}_{fin}])$ and $\Cb(\nu_{in};[\la_{in},{\la}_{fin}])$.
\smallskip
 
 d) The costraint equations {\em (I;1.55),(I;1.56)} are the pull back of (some of) the structure equations of ${\cal K}({\overline\tau})$
restricted to $C(\la_{in};[\nu_{in},{\nu}_{fin}])$ and $\Cb(\nu_{in};[\la_{in},{\la}_{fin}])$.
\smallskip

 e)  It is possible to  define on ${\cal K}({\overline\tau})$ a null orthonormal frame $\{e_4,e_3,e_a\}$ adapted to the ``double null
canonical foliation", the metric components, $\ga_{ab},\oom,X^a$ with respect to the adapted coordinates
$\{u,\ub,\theta,\phi\}$ and the corresponding connection coefficients
\bea
\chi_{ab}&=&\ggg(\dd_{e_a}e_4,e_b)\ \ ,\ \chib_{ab}=\ggg(\dd_{e_a}e_3,e_b)\nn\\
\omb&=&-\frac{1}{2}\dd_3(\log\oom)\ ,\ \om=-\frac{1}{2}\dd_4(\log\oom)\\
\ze_{a}&=&\frac{1}{2}\ggg(\dd_{e_{a}}e_4,e_3)\ .\nn
\eea
Moreover  a vector field  $X^a$ can be defined in ${\cal K}({\overline\tau})$ such that denoting $N$ and $\Nb$ two null vector fields equivariant
with respect to the $S(\la,\nu)$ surfaces, the following relations hold: $N=\oom e_4$, $\Nb=\oom e_3$ and, in the $\{u,\ub,\theta,\phi\}$
coordinates, 
\[N=2\oom^2\frac{\partial}{\partial\ub}\ ,\ \Nb=2\oom^2\left(\frac{\partial}{\partial u}+X^a\frac{\partial}{\partial\om^a}\right)\ .\]
\end{theorem}
{\bf Remarks:} 

a) The structure of the local existence theorem which we will use repeatedly in the following, is the
same as that of the first part of the ``characteristic global existence theorem", Theorem \ref{T2.1}. The basic difference is that in 
Theorem \ref{T2.1}, the region ${\cal K}(\tau_*)$ is the maximal development of the initial data characteristic
hypersurface $C(\la_1;[\nu_0,\nu_1])\cup\Cb(\nu_0;[\la_1,\la_0])$ and $\tau_*$ is not small.
\smallskip

b) The subsets of the regions $\Delta_1{\cal K},\Delta_2{\cal K},\Delta_3{\cal K}$ whose existence is proved by the local existence
result, correspond to different values of $\la_{in}$ and $\nu_{in}$, namely \footnote{The final value $\de$ will be the smallest between
$\de'$,$\de''$,$\de'''$.}
\beaa
&&\Delta^{\!loc}_1{\cal K}: \la_{in}=\la_1\ ,\ \nu_{in}=\overline\nu\ ,\ \la_{fin}=\la_{in}-\de'\ ,\ \nu_{fin}=\nu_{in}+\de'\nn\\
&&\Delta^{\!loc}_2{\cal K}: \la_{in}=\la_0\ ,\ \nu_{in}=\nu_0\ ,\ \la_{fin}=\la_{in}-\de''\ ,\ \nu_{fin}=\nu_{in}+\de''\nn\\
&&\Delta^{\!loc}_3{\cal K}: \la_{in}={\overline\la}(\nu_*)\ ,\ \nu_{in}=\nu_*\ ,\ \la_{fin}=\la_{in}-\de'''\ ,\
\nu_{fin}=\nu_{in}+\de'''\ \ .\nn
\eeaa

c) The initial data for the region $\Delta_1{\cal K}$ are given on the initial null hypersurfaces
$C(\la_1;[\nu_*,\nu_*+\de])\cup\Cb(\nu_*;[\la_1,\la_0])$ therefore a part of them are (a portion of) the initial data of the initial problem, those
on $\Cb(\nu_*;[\la_1,\la_0])$ are provided by the values that the various connection coefficients and their derivatives acquire on the
upper boundary of the region ${\cal K}(\tau_*)$. Also these data are small as follows from the control of the $\cal O$ norms in this region, see
subsection \ref{SS3.2}. Moreover as the initial data are expressed in terms of the connection coefficients, see (I),
they automatically satisfy the constraints equations of the characteristic problem.

\NI Exactly the same argument holds for $\Delta_2{\cal K}$, while for $\Delta_3{\cal K}$ the initial data for extending our result to
this region are obtained after the construction of regions $\Delta_1{\cal K}$ and
$\Delta_2{\cal K}$ is completed.
\smallskip

d) The proof of the local existence result will not be given in this paper as, basically, it already exists due to A.Rendall,
\cite{rendall:charact}, D.Christodoulou and H.Muller Zum Hagen, \cite{ChristMuller}, \cite{Muller}. Nevertheless as their proof is given
in the harmonic gauge we have to connect the initial data norms we use in this paper with those defined in the harmonic gauge. This
connection has been discussed at length in section 4 of (I) and we recall some aspects of it in Section \ref{S5}. There we also
examine the local existence proof in \cite{ChristMuller}, \cite{Muller} and give some hints for a different proof using a
Cauchy-Kowalevski type argument, \cite{NicCK}.

\subsection{Better norm estimates in ${\cal K}(\tau_*)$}\label{SS3.2}
This is the first step required to prove our result. This is also the more difficult one, but in the present case is a
repetition of part of the proof given in \cite{Kl-Ni:book}, Chapters 4-5-6. The main difference is in the nature of the
initial data given here on two null hypersurfaces. Nevertheless even when the data are assigned, proving the result presents some
differences and some simplifications.
\smallskip

a) Due to the nature of the initial hypersurfaces, a problem present in \cite{Kl-Ni:book} is not present. There, in fact, once a
double null canonical foliation was obtained, it was possible to define a global time function, $t(p)=\frac{1}{2}(u(p)+\ub(p))$ and prove that
the two dimensional surfaces $S(\la,\nu)=C(\la)\cap\Cb(\nu)$ were embedded in the spacelike hypersurfaces, $\Si_t$, level surfaces of the time
function. The problem was that these spacelike hypersurfaces were not maximal and, as $t\rightarrow 0$, they did not coincide with the
(spacelike) initial data hypersurface. This required a delicate control of a small slab near $\Si_0$ to show how the initial data could be
transported along the outgoing and incoming cones, See \cite{Kl-Ni:book} section 4.1.3 . In the present case this problem is completely avoided
as the two initial data null hypersurfaces belong to the double null canonical foliation.
\smallskip

b) The second difference, but not a simplification, is that, as discussed in (I, subsubsection 3.1.3), a greater regularity
for the initial data along the directions ``$S$-tangential"\footnote{The directions tangential to the two dimensional surfaces $S(\la_1,\nu)$ and
$S(\la,\nu_0)$.} to the initial null hypersurfaces, is needed to have a sufficient control of the initial data derivatives in the direction
``orthogonal" to them. This higher regularity is needed only for the initial data and, contrarily to what said in (I), a control of higher
derivatives in the region ${\cal K}(\nu_*)$ is not needed.


\NI This discussion can be summarized in the following theorem:
\begin{theorem}\label{Th3.2}
Let $({\cal K}(\tau),\ggg)$ be a solution of the ``characteristic
Cauchy problem" relative to initial data \footnote{More precisely the portion of initial data on
$C(\la_1;[\nu_0,{\overline\nu}])\cup\Cb(\nu_0;[\la_1,{\overline\la}])$.} satisfying
$J^{(q)}_{C_0\cup\Cb_0}\leq\varepsilon$ and inequalities
\ref{2.3rt}, with the following properties:

{\bf i)} $({\cal K}(\tau),\ggg)$ is foliated by a double null canonical foliation $\{C(\la)\}$, $\{\Cb(\nu)\}$
with $\la\in [\la_1,{\overline\la}]\ ,\ \nu\in [\nu_0,{\overline\nu}]$. Moreover
\beaa 
&&i(C_0)\cap{\cal K}(\tau)=C(\la_1;[\nu_0,{\overline\nu}])\ ;\  i(\Cb_0)\cap{\cal K}(\tau)=\Cb(\nu_0;[\la_1,{\overline\la}])
\nn\\
&& where \  \  \ \ \ \ \ \  \ \ \ \ \ \ {\overline\nu}+{\overline\la}=2\tau\ \ \ \ \ ;\ \ \ \ \
{\overline\nu}-{\overline\la}=2\ro_0\ . \ \ \ \ \ \  \ \ \ \ \ \  \ \ \ \ \ \  \ \ \ \ \ \  \ \ \ \ \ \  \ \ \ \ \ \  \ \ \ \ \ \  
\eeaa

{\bf ii)}
Let $\{e_3,e_4,e_a\}$ be a null orthonormal frame adapted to the double null canonical foliation. Let ${\cal R},{\cal O}$ be the family of norms
for the null components of the Riemann curvature tensor and for the connection coefficients relative to this null frame, given in
\cite{Kl-Ni:book}, Chapter 3.  Given $\ep_0>0$ sufficiently small, but larger than $\varepsilon$, assume that the norms ${\cal R}, {\cal O}$
satisfy the following inequalities
\bea
{\cal R}\leq \ep_0\ \ ,\ \ {\cal O}\leq \ep_0\ .\eql{3.9z}
\eea

{\bf iii)} Denoted by $\cal T$ the set of all values $\tau$ for which the spacetime
${\cal K}(\tau)$ does exist, we define $\tau_*$ as the $\sup$ over all the values of $\tau\in{\cal T}$:
\bea
\tau_* =\sup \{\tau\in {\cal T}\}\eql{2.58z}\ .
\eea
Under all these assumptions it follows that in ${\cal K}(\tau)$ these norms satisfy the bounds
\bea
{\cal R}\leq c\varepsilon\ \ ,\ \ {\cal O}\leq c\varepsilon\ ,
\eea
an improvement of \ref{3.9z}, if $\varepsilon$ is chosen sufficiently small.
\end{theorem}
{\bf Proof:} The theorem is a repetition of a set of theorems proved in \cite{Kl-Ni:book}, more specifically Theorems {M1}, {M2},
{M3}, {M4}, {M5}, {M7}, {M8}. The main difference is that the norms $\cal O$ and $\cal R$ relative to the initial data, contains more
$\nabb$-derivatives than those required in \cite{Kl-Ni:book}. We specify the chosen order of derivatives for the initial data norms
in Section \ref{SS3.4}. 

\subsection[Existence proof for the strips $\Delta_1{\cal K}$, $\Delta_2{\cal K}$\label{SS3.4}
and the diamond $\Delta_3{\cal K}$ ]{Existence proof for the strips $\Delta_1{\cal K}$, $\Delta_2{\cal K}$
and the diamond $\Delta_3{\cal K}$}\label{SS.3.4}
\subsubsection{$\Delta_1{\cal K}$\ ,\ $\Delta_2{\cal K}$}\label{SSS.3.4.1}

The first observation is that proving the existence of the strip $\Delta_1{\cal K}$ amounts to prove again a ``global" existence result. In fact
this strip, although very narrow in the outgoing future direction ($\de$ can be chosen as small as we like), has a length, measured
with the affine parameter $\la$, which can be arbitrarily large. Recall that the ``length" of $\Delta_1{\cal K}$,
$\ |{\overline\la}(\nu_*)-\la_1|$,  depends linearily on $\nu_*$, see equation \ref{2.19y}, and that $\nu_*$ can be arbitrary large
depending on the choice of
$\ro_0$, see equation \ref{3.22ggw}.

\NI This is a somewhat delicate point. First of all we observe that the existence proof for the strip $\Delta_1{\cal K}$ is significantly
different from the analogous step in the non characteristic result of \cite{C-K:book}. In fact in that case the spacetime is foliated by
spacelike hypersurfaces $\{\Si_t\}$ and going from $\Si_{t_*}$ to $\Si_{t_*+\de}$ is a purely local existence result while in the present case
the strip $\Delta_1{\cal K}$ extends toward the future (in time) for a possibly very large interval. This could seem, at first sight, a
serious problem as to prove the existence of the maximal Cauchy development associated to the initial data, that is of the unbounded
spacetime region with boundary $C(\la_1)\cup\Cb(\nu_0;[\la_1,\la_0])\cup C(\la_0)$, we need to prove preliminary the existence of the
(arbitrary large) strip region $\Delta_1{\cal K}$ which looks as        a region of the same type.

\NI The  contradiction is only apparent due to the fact that the strip $\Delta_1{\cal K}$ is arbitrarily narrow (of order
$\de$) which allows to prove its existence in an easier way, while the width of our final region\footnote{Either if  we consider the first or the
second part of our global result.} is of order ``$|\la_1-\la_0|$" which can be arbitrarily large.
 
\NI To prove the existence of $\Delta_1{\cal K}$ we prove, first, via the local existence theorem, that at least a portion,
$\Delta^{\!loc}_1{\cal K}$, of $\Delta_1{\cal K}$ exists whose norms satisfy, see figure 7,
\bea
{\cal O}\leq \ep_0\ \ ,\ \ {\cal R}\leq \ep_0\ \ \ .\eql{5.1az}
\eea
Second we define $\la_*$ as the sup of all the values $\la\in[\la_1,\la_0]$ such
that the portion $\Delta_1{\cal K}(\la)$ of the strip $\Delta_1{\cal K}$ exists satisfying \ref{5.1az}.
We denote this region $\Delta_1{\cal K}(\la_*)$.

\NI Third, using the fact that on $C(\la_1,[\nu_*,\nu_*+\de])\cup\Cb(\nu_*;[\la_1,\la_*])$, the lower part of the boundary of $\Delta_1{\cal
K}$, the data are bounded by $c\varepsilon$, we show that in $\Delta_1{\cal K}(\la_*)$ the norms $\cal O$ and $\cal R$ are bounded by
$C\varepsilon$. This allows to extend, again via the local existence theorem, the region $\Delta_1{\cal K}(\la_*)$ in the $e_3$
direction and prove that, therefore, $\la_*$ cannot be the $\sup$ for all the values $\la\in [\la_1,\la_0]$. To avoid a contradiction,
therefore, the whole strip $\Delta_1{\cal K}\equiv\Delta_1{\cal K}(\la_0)$ must exist.

\NI As we said this proof is an easier version of the general one, but the first step is analogous to
the one discussed in subsection \ref{SS3.2} for the general case, namely we prove that in $\Delta_1{\cal K}(\la_*)$ the following
inequalities hold
\bea
{\cal O}\leq C\varepsilon\ \ ,\ \ {\cal R}\leq C\varepsilon\ .\eql{5.1a}
\eea
This result is proven with the same strategy used in \cite{Kl-Ni:book}: assuming first that ${\cal O}\leq\ep_0$ we prove that
the generalized energy norms $\cal Q$ can be controlled by the same norms on the ``initial" boundary of
$\Delta_1{\cal K}(\la_*)$ and therefore  bounded by $C\varepsilon$; from these estimates ${\cal R}\leq C\varepsilon$ follows.
Using this result and the transport equations we show that ${\cal O}\leq C\varepsilon$ holds.

\NI To understand where the narrowness of $\Delta_1{\cal K}$ is used let us examine in more detail how the first part of this result is
obtained namely the proof of the following lemma
\begin{Le}\label{Le3.1}
Assume that $\Delta_1{\cal K}(\la_*)$ be foliated by a double null foliation made by null hypersurfaces which are the level hypersurfaces of two
functions $u(p)$ and $\ub(p)$ solutions of the eikonal equation with initial data on $\Cb(\nu_*;[\la_1,\la_*])$ and $C(\la_1;[\nu_*,\nu_*+\de])$
respectively, given by two functions $u|_{\Cb(\nu_*;[\la_1,\la_*])}(p)$ and $\ub_{C(\la_1;[\nu_*,\nu_*+\de])}(p)$ which define on these null
hypersurfaces a canonical foliation as discussed in subsection 3.3.1 of {\em\cite{Kl-Ni:book}}, see also {\em\cite{Niclast}}. Then if in
$\Delta_1{\cal K}(\la_*)$ the Riemann norms satisfy ${\cal R}\leq c\varepsilon$ it follows that,
\[{\cal O}\leq C\varepsilon\ .\] 
\end{Le}
{\bf Remark:} To prove this result we have only to show that the region $\Delta_1{\cal K}(\la_*)$  can be large in the incoming
direction, that is for values of $\la>\la_1$, while in the outgoing direction as the interval $[\nu_*,\nu_*+\de]$, can be chosen
arbitrarily small, we do not need an extension in that direction. In fact $\de$ can be the variation in $\nu$ obtained in the
local existence. This implies that all the norm of the not underlined connection coefficients collected in $\cal O$,\footnote{recall that in
$\cal O$ there are also the norms of the underlined connection coefficients.} can be estimated as in Chapter 4 of
\cite{Kl-Ni:book}, but integrating, on the outgoing cones, toward the future direction. This fact, possible for the
smallness of $\de$, makes the proof of the existence of $\Delta_1{\cal K}$ a different result than the global one we are looking for. In
other words a bootstrap argument is required here only for the $e_3$ ($\la$-direction) and not for the $e_4$ ($\nu$-direction).
\smallskip

\NI {\bf Proof:} To control the norms of the connection coefficients and their derivatives collected in $\cal O$ we have to examine their transport
equations along the incoming and outgoing (portions of) cones contained in $\Delta_1{\cal K}(\la_*)$. Those associated to the underlined
connection coefficients, along the incoming cones, are exactly of the same type as the corresponding ones in Chapter 4
of \cite{Kl-Ni:book}. Those associated to the (``small" portions of) outgoing cones allow to obtain the norms of the non underlined connection
coefficients as an integral along the portions of the outgoing cones plus their initial value on $\Cb(\nu_*;[\la_1,\la_*])$.
These last estimates differ from the corresponding ones of Chapter 4 of \cite{Kl-Ni:book} where the integrations are done, using the transport
equations along the ``outgoing cones", starting from above and going down backward in time.

\NI The choice of integrating along the outgoing ``cones" from the above is necessary in the global existence proof of
\cite{Kl-Ni:book} to obtain the right decay factors, but in the present situation, due to the smallness od $\de$ we can, and have to, proceed
in the opposite way. 

\NI To clarify this point let us recall the logic for the estimates of the $\cal O$ and $\cal R$ norms used in \cite{Kl-Ni:book}. This is
basically the content of Chapter 4; the main assumption is that there exists a spacetime region where $\cal O$ and $\cal R$ are bounded
by $\ep_0$. This assumption done, one proves that $\cal O$ and $\cal R$ are in fact bounded by $C\varepsilon$ where $\varepsilon$ indicates
the smallness of initial data and can be chosen sufficiently small so that         $C\varepsilon\!\leq\!\frac{1}{2}\ep_0$. The estimates for
$\cal R$ are obtained once we control the $\cal Q$ norms in terms of the same norms on the ``initial" boundary of $\Delta_1{\cal K}(\la_*)$
and the assumption ${\cal O}\leq\ep_0$. The estimates for $\cal O$ are obtained using the bound ${\cal R}\leq C\varepsilon$ the transport
equations and the estimates ${\cal O}\leq C\varepsilon$ on the
$\Cb(\nu_*;[\la_1,\la_*])$ part of the $\Delta_1{\cal K}(\la_*)$ boundary. 

The main difference with the general case is in the estimate for $\cal Q$. This is achieved through a volume integration on the whole
region which allows to control the $\cal Q$ integrals on the ``upper" boundaries in terms of those along the
``lower" ones plus some volume integrals denoted  ``error terms". Estimating the ``error terms", one can show
that they are $O(\ep_0^2)$, using the previous assumptions on $\cal O$ and $\cal R$, therefore showing that the integral norms $\cal Q$ can
be bounded in terms of the same norms made with the initial data and, therefore, by $C\varepsilon$. In the \cite{Kl-Ni:book} two conditions
are required to prove these steps: the correct asymptotic decay and the correct regularity of the various connection
coefficients.\footnote{The second requirement arises from the fact that the the connection coefficients have to be controlled up to third
order, for controlling the Riemann tensor up to second order; this implies that to achieve the result the estimate of the $n$-order
derivatives of the connection coefficients must require only the control of the $n-1$-order for the Riemann tensor. This is the reason why we
need the so called ``last canonical foliation" procedure, see the detailed dicussion in \cite{Kl-Ni:book}.} As said before, in the present
case, due to the fact that the
$\Delta_{1}{\cal K}$ regions are ``narrow" the decay is not important in the outgoing direction and we can integrate from below. This
has also the (crucial) side effect that, as on the ``lower" boundary of $\Delta_{1}{\cal K}$ a canonical foliation has already been
established, the needed regularity of the various connection coefficients and Riemann components is immediately
obtained.\footnote{If this were not the case we should have to assume that on the ``upper boundary" of $\Delta_{1}{\cal K}$ a
canonical foliation is given, but this requires having on it already a background foliation with the appropriate estimates for $\cal
R$ and $\cal O$ ($\leq\ep_0$). These estimates cannot be obtained if the region is not narrow due to the lack of the right
regularities relations between the $\cal O$ coefficients and the $\cal R$ components. A canonical foliation on the ``last slice" could
be assumed as in the general proof, but then one should again prove a bootstrap argument also in the $\nu$ direction requiring again a
semilocal extension and so on and so for.}
The discussion for $\Delta_{2}{\cal K}$ is exactly the same with the obvious interchanges and we do not repeat it here.
\subsubsection{$\Delta_3{\cal K}$}\label{SSS.3.4.2}
The existence proof for the region $\Delta_3{\cal K}$ is simpler as it is strictly a local existence proof. In fact this region is the
maximal development of the null surface $C({\overline\la}(\nu_*);[\nu_*,\nu_*+\de])\cup\Cb(\nu_*;[{\overline\la}(\nu_*),{\overline\la}(\nu_*+\de)])$
which can be made as small as we like just decreasing the parameter $\de$. Again on this surface one has to build a canonical foliation exactly
following the strategy described in \cite{Kl-Ni:book} and \cite{Niclast}.

\subsection{The regularity of the initial data}\label{SS3.4}
In this subsection we determine the minimum regularity of the initial data required to achieve our existence proof.
Here the situation is more complicated than in the non characteristic case, studied in \cite{C-K:book} and \cite{Kl-Ni:book}, due to
the presence of the ``loss of derivatives" phenomenon discussed in (I), see also \cite{Muller} for the local existence result,  which
implies that a greater regularity is needed. 

\NI The basic fact to determine the due regularity for the initial data is that to prove the global existence result
 we have to control in the $L^2(C)$ norms,\footnote{Here $C$ is a null incoming or outgoing cone of the double null canonical foliation.} in the
bulk region, up to second derivatives of the Riemann tensor and third derivatives of the connection connections, in other words up to $s=4$
derivatives of the metric components.

\NI Fixed $s$ for the spacetime Sobolev norms, the local existence theorem or more precisely the costraints on the initial data for the
characteristic local existence, see \cite{Muller}, imply that the angular derivatives of the initial data metric components must be controlled
up to order $2s-1$. The same result we obtain here as we are going to discuss in some detail.

\NI In subsection 3.1.1 of (I) we showed how the $C^{\a}$ regularities of the various connection coefficients
and the null Riemann components are connected; to adapt it to the Sobolev case we proceed as follows:

\NI From (I) we have on $C_0$
\bea
&&\oom,\ \ga,\ \chi,\ \om,\ \dddd_4\chih\in C^{q'}(S)\nn\\
&&X,\ \ze,\ \nabb\log\oom\in C^{q'-1}(S)\eql{4.1qw}\\
&& \chib,\ \omb \in C^{q'-2}(S)\ \ .\nn
\eea 
and from the regularity of the metric components and the connection coefficients we have immediately, see (I) for more details,
\bea
\a\in C^{q'}(S)\ ,\ \b\in C^{q'-1}(S)\ ,\ (\ro,\si)\in C^{q'-2}(S)\ ,\ \bb\in C^{q'-3}(S)\ .\eql{4.2qw}
\eea
As we will impose the needed regularity in terms of the $H^s_p(S)$ norms, \footnote{Recall that $\sup_S|G|\leq
cr^{-\frac{1}{2}}\|G\|_{H^1_4(S)}$, see \cite{Kl-Ni:book}, Lemma 4.1.3.} conditions
\ref{4.1qw},
\ref{4.2qw} can be reexpressed in the following way, with $p\in[2,4]$, 
\bea
&&\oom,\ \ga,\ \chi,\ \om,\ \dddd_4\chih\in H_p^{q'+1}(S)\nn\\
&&X,\ \ze,\ \nabb\log\oom\in H_p^{q'}(S)\ \ ,\ \ \chib,\ \omb \in H_p^{q'-1}(S)\eql{4.1qwz}\\
&&\a\in H_p^{q'+1}(S)\ ,\ \b\in H_p^{q'}(S)\ ,\ (\ro,\si)\in H_p^{q'-1}(S)\ ,\ \bb\in H_p^{q'-2}(S)\ .\nn
\eea
To determine the minimum value of $q'$ let us consider the connection coefficient $\chibh$ and observe that from the
explicit expression of $\bb$, 
\[\bb=-[\nabb\tr\chib-\divv\chib+\zeta\c\chib-\zeta\tr\chib]\ ,\]
it follows that $\bb=O(\nabb\chibh)+other\ terms$. From the explicit expression of the norms $\cal R$, see the appendix, it follows that on
the spacetime and, therefore, on $C_0$ the following norm has to be bounded and small:
\bea
\|\ddb_3^2\bb\|_{L^2(C_0)}=O(\ep_0)\ \ \mbox{which implies}\ \  \|\ddb_3^2\nabb\chibh\|_{L^2(C_0)}=O(\ep_0)\ .
\eea
Due to the loss of derivatives this implies, at its turn, that on $C_0$ we must have also
\bea
\|\nabb^5\chibh\|_{L^2(C_0)}=O(\ep_0)\ .\eql{3.17q}
\eea
As we impose the initial conditions in terms of the $H^s_p(S)$ norms, \ref{3.17q} is satisfied requiring
\bea
\|\nabb^5\chibh\|_{H_p^0(S)}=O(\ep_0)\  \mbox{which also implies}\ \|\chibh\|_{H_p^5(S)}=O(\ep_0)\ ,\eql{3.18q}
\eea
where we always ignored the appropriate weights, $r^{(\a-\frac{2}{p})}$, of the norms. Condition \ref{3.18q} compared with \ref{4.1qwz} gives
$q'-1=5$\ {and therefore}  
\bea
&&\oom,\ \ga,\ \chi,\ \om,\ \dddd_4\chih\in H_p^{7}(S)\nn\\
&&X,\ \ze,\ \nabb\log\oom\in H_p^{6}(S)\ \ ,\ \ \chib,\ \omb \in H_p^{5}(S)\eql{4.1qwzz}\\
&&\a\in H_p^{7}(S)\ ,\ \b\in H_p^{6}(S)\ ,\ (\ro,\si)\in H_p^{5}(S)\ ,\ \bb\in H_p^{4}(S)\ .\nn
\eea
In conclusion the minimum value of $q$ in $J^{(q)}_{C_0\cup\Cb_0}$, see \ref{2.2w}, is $q=7$.

\NI{\bf Remarks:}
\smallskip

a) Observe that the requirement $q=7$ is connected, as we said, to the characteristic constraint equations which implies the so called
``loss of derivatives" and also to the Bianchi equations. In fact let us consider the null Riemann component
\[\a=-[\dddd_4\chih+\tr\chi\chih-(\dd_4\log\oom)\chih]\ ;\]
From the explicit expression of $\cal R$ it follows that we have to require that
\beaa
\|\dddd_3^2\a\|_{L^2(C_0)}=O(\ep_0)\ \ \mbox{implying, proceeding as before,}\ \ \|\nabb^4\chih\|_{L^2(C_0)}=O(\ep_0)\ .
\eeaa
Again this is satisfied imposing
\bea
\|\nabb^4\chih\|_{H^0_p(S)}=O(\ep_0)\ \ \mbox{which requires}\ \ \|\chih\|_{H^4_p(S)}=O(\ep_0)\ .
\eea
To satisfy this bound it would be enough to choose   $q'=3$, implying $q=4$ instead of $q=7$. Nevertheless this would not be sufficient as the
Bianchi equations in the outgoing direction imply that if $\a\in H^4_p(S)$ then $\bb\in H^1_p(S)$, while we know that we have to
control $\|\bb\|^4_p(S)$.
\smallskip

b) The function of the initial data $J^{(q)}_{C_0\cup\Cb_0}$ which we impose to be small, is a function of the $|\c|_{p,S}$ norms.
Nevertheless the initial data smallness could also be prescribed in terms of the $H^s(C)$ norms. We could also in this case require $s=7$
and this would imply less regularity for the initial data, but we will not discuss this aspect any further here.
\section[The last slice canonical foliation]{The existence proof for the last slice canonical foliation}\label{S4}
In this section we discuss in which part of the boundaries of the different regions a canonical foliation has to be imposed. This has been
claimed many times and we make now these statements more precise.

\NI We do not give here the definition of a canonical foliation of a null hypersurface; the reader can find it,  in any detail, in
\cite{Kl-Ni:book}, Chapter 4, and in \cite{Niclast}. Either we do not discuss why these foliations are needed, but we only recall
that they allow to control the estimates of the $n+1$ derivatives of the connection coefficients in terms of the estimates for the $n$
derivatives of the Riemann tensor, see Lemma \ref{Le3.1} and footnote 22. This is a crucial step for the implementation of the
bootstrap argument, proving Theorem \ref{Th3.2}.
\smallskip

\NI That (part of) the boundaries of the various regions have to be canonically foliated is required in various occasions, more
precisely:
\smallskip

i) In the definition of $\cal{K}(\tau)$ the portion $C(\la_1;[\nu_0,\overline{\nu}])\cup\Cb(\overline{\nu};[\la_1,\overline{\la}])$ of the
boundary has to be foliated with a canonical foliation. 
\smallskip

ii) The proof that the region $\cal{K}(\tau_*)$ can be extended to $\cal{K}(\tau_*+\de)$, requires that this extension has the same properties
of  $\cal{K}(\tau)$. 
Recalling that 
\bea
{\cal{K}(\tau_*+\de)}={\cal K}(\tau_*)\cup\Delta_1{\cal K}\cup\Delta_2{\cal K}\cup\Delta_3{\cal K} 
\eea
this implies that part of the boundary of $\Delta_1{\cal K}$ has to be canonically foliated. Observe that
\bea
\partial\Delta_1{\cal K}\!&=&\!C(\la_1;[\nu_*,\nu_*+\de])\cup\Cb({\nu_*};[\la_1,\overline{\la}(\nu_*)])
\cup C(\overline{\la}(\nu_*);[\nu_*,\nu_*+\de])\nn\\
&&\cup\Cb(\nu_*+\de;[\la_1,\overline{\la}(\nu_*)])\ ,
\eea
therefore to make the spacetime ${\cal{K}(\tau_*+\de)}$ fulfilling all the requested conditions, the portion
$C(\la_1;[\nu_*,\nu_*+\de])\cup\Cb(\nu_*+\de;[\la_1,\overline{\la}(\nu_*)])$  of $\partial\Delta_1{\cal K}$ has to be canonically
foliated. Remark that, nevertheless, to prove the existence of $\Delta_1{\cal K}$ it is enough to have a canonical foliation on
$C(\la_1;[\nu_*,\nu_*+\de])$.\footnote{As for the smallness of $\de$ in $\Delta_1{\cal K}$ we can integrate toward the future in the outgoing
direction, it is enough that a canonical foliation be present on $\Cb(\nu_*;[\la_1,\overline{\la}(\nu_*)])$.}
The canonical foliation on    $\Cb(\nu_*+\de;[\la_1,\overline{\la}(\nu_*)])$ has to be built after the region $\Delta_1{\cal K}$ has been
obtained and it is only required to have the region ${\cal{K}(\tau_*+\de)}$ with the right properties, i),..., vi), of Section \ref{S2}.

\NI The situation is quite similar for the region $\Delta_2{\cal K}$; to prove its existence the only part of the boundary which has to be
canonically foliated is $\Cb(\nu_*;[\overline{\la}(\nu_*),\overline{\la}(\nu_*)+\de])$. In fact it is clear that on
$C(\overline{\la}(\nu_*);[\nu_0,\nu_*])$ the foliation induced from the incoming ``cones" starting from the canonical foliation defined on
$C(\la_1;[\nu_0,\nu_*])$ is already canonical.\footnote{A slight difference in the ``semilocal" existence proof for this region is that in
this case the assumptions for the regions $\Delta_2{\cal K}(\nu)$ require that the portion of its boundary,
$\Cb(\nu;[\overline{\la}(\nu),\overline{\la}(\nu)+\de])$, has to be canonically foliated.}

\NI Finally the existence proof of the region $\Delta_3{\cal K}$ is only a local problem. Therefore it requires only that
the lower part of its boundary has the right properties. For $C(\overline{\la}(\nu_*);[\nu_*,\nu_*+\de])$ this follows from the
canonical foliation of $C(\la_1;[\nu_*,\nu_*+\de])$ and for $\Cb(\nu_*;[\overline{\la}(\nu_*),\overline{\la}(\nu_*)+\de])$ from the requirement
which has been satisfied in the construction of $\Delta_2{\cal K}$. Again, once $\Delta_3{\cal K}$ has been obtained, a canonical
foliation has to be imposed on $\Cb(\nu_*+\de;[\overline{\la}(\nu_*),\overline{\la}(\nu_*)+\de])$.\footnote{Once
$\Delta_1{\cal K}$ and $\Delta_3{\cal K}$ are obtained, the procedure will be to impose a canonical foliation on the
whole $\Cb(\nu_*+\de;[{\la}_1,\overline{\la}(\nu_*)+\de])$.} 

\NI This completes the description of how the canonical foliation has to be placed on the various null hypersurfaces. The existence proof of
these foliations is in \cite{Niclast} and looking at the various steps of the construction of the various regions involved and to the initial
data conditions it is easy to realize that the assumptions required in this proof are always satisfied.
\smallskip

\NI {\bf Remark:} As already said, imposing on the boundary a canonical foliation is done for being able to control, in the spacetime, the
$n+1$ derivatives of the connection coefficients in terms of the $n$ derivatives of the Riemann tensor, avoiding any loss of derivatives in
the spacetime. As, on the other side, the loss of derivatives on the initial hypersurfaces, due to their characteristic nature, is
inavoidable and force us to require a greater regularity on $C_0\cup\Cb_0$ than the one needed in the spacetime interior, it follows that, in
principle, we could avoid the request of having a canonical foliation on $C_0$. 
\section{The local existence result  }\label{S5}
In this section we give some hints about the local existence result we need to complete our global proof and which has
been stated in Section \ref{S.3}, Theorem \ref{theor3.1}.

\NI We do not present here the details of the proof, we describe only two different procedures to achieve the result. The reason
is twofold, first there are in the literature various proofs basically fulfilling our goal, in particular we refer to those of A.Rendall,
\cite{rendall:charact}, H.Muller Zum Hagen, \cite{Muller} and M.Dossa, \cite{Dossa}; second in a subsequent paper one of the authors (F.N.) will
present a new proof of the local existence result based on a Cauchy-Kowalevski type argument, \cite{NicCK}.
\smallskip

\NI Assume we want to use the work of A.Rendall, \cite{rendall:charact} or the one of H.Muller Zum Hagen
\cite{Muller}. The main difficulty in adapting these results to our case is that they have been obtained using harmonic coordinates while we
want that the ``gauge" for ${\cal K}(\tau)$ be the one associated to the double null canonical foliation, we called in (I) ``$\oom$-gauge". This
requires to state precisely the connection between the initial data written in these two different gauges so that we can reexpress their results
in our formalism. 
\subsubsection{The relation between the harmonic and the $\Omega$-foliation gauges.}
This has been discussed in great detail in (I), here we just recall the rational behind it. In (I) the initial data are given on two null
hypersurfaces foliated by two dimensional surfaces $S^2$-diffeomorphic. They are expressed in terms of connection coefficients which are tensors on
the leaves of these foliations and of the metric defined on them. The way they have to be assigned, to satisfy the costraints, has been discussed in
(I). On the other side as the local existence result has been proved in the harmonic gauge, we have to express our data in this gauge. This is
the content of subsection 4.1 of (I). Once this has been done we are allowed to use the harmonic local existence result.

\NI Nevertheless the goal in not yet achieved as in the local result obtained, for instance in \cite{Muller}, the region whose existence is
proved is foliated by spacelike hypersurfaces defined as level hypersurfaces of a time function $\tau(p)$ which, if the initial data are
``small", approximate the portion of $t$-constant hyperplanes intersecting the region inside the null initial hypersufaces.
Therefore to obtain the local region we need to complete this result adding to this region the local region which is the
maximal development of the ``data" on the last $\tau$-leave of the previous local result. In this way a small ``diamond"-shaped region
is built and, finally, the proof that in this region the norms $\cal R$ and $\cal O$ are bounded by $\ep_0$ goes as discussed before.

\NI The construction of the local solution envisaged now can be avoided just trying to get a local solution proceeding in a more straightforward way
from the Einstein equations written in the ``$\oom$-gauge". The basic idea is to write those structure equations for
the connection coefficients which are evolution equations along a null direction, for instance the incoming one, see equations (1.8), (1.10),
(1.17), (1.18), (1.20) of (I), together with the equations defining them in terms of the derivatives of the metric tensor $\ga_{ab}$. The solutions of
this set of equations satisfying also the set of equations associated to the structure equations along the outgoing direction  are solutions of the
vacuum Einstein equations. One could, therefore, obtain  a real analytic solution applying the Cauchy-Kowalevski result to the characteristic
case, see \cite{Duff}, and then look for a less regular ($H_p^s$) solution as a limit of a sequence of real analytic solutions. Here,
nevertheless, a problem arises: to build   a sequence of solutions converging in an appropriate norm once the initial data converge to less
regular data, we have to use energy-type norms and prove that they are bounded. Using these norms one can show the convergence of a sequence of
analytic real solutions. The problem is associated to the fact that the Einstein equations written in the ``$\Omega$-gauge" are not of
hyperbolic type and therefore we do not have obvious energy norms at our disposal like those used in the harmonic gauge. To overcome this
difficulty we mimic the strategy of \cite{C-K:book} and
\cite{Kl-Ni:book} where the energy-type norms used are those made with the Bel-Robinson tensor.
We can solve, therefore, this set of equations, applying Cauchy-Kowalevski theorem and then look for a sequence
of solutions converging in the appropriate ``Bel-Robinson"-energy norms to the (local) solution we are looking for.
\section[Appendix]{Appendix}

\subsection{The $\cal O$ and the $\cal R$ norms}\label{SSA.1} 

Due to the fact that in the characteristic case the regularity required to the initial data is greater than the regularity achieved from the solution
we write here the norms which are used for the initial data.

\subsubsection{The initial data norms}\label{SSSA.1} 
The initial data norms are regrouped in two sums, one,
$J^{(q)}_{C_0}\left[\overline{\ga}_{ab},\overline{\oom},\overline{\ze}_a,\overline{\chib},\overline{\omb}\right]$, appropriate for the outgoing cone
and the second, $J^{(q)}_{\Cb_0}\left[\overline{\ga}_{ab},\overline{\oomb},\overline{\underline
X}^a,\overline{\ze}_a,\overline{\chi},\overline{\om}\right]$, for the incoming one.
Their explicit expressions were given in (I) and we repeat them here, correcting some misprints,

\bea
&&J^{(q)}_{C_0}\left[\overline{\ga}_{ab},\overline{\oom},\overline{\ze}_a,\overline{\chib},\overline{\omb}\right]=
\sup_{C_0}\left(r\big|\overline{\oom}-\frac{1}{2}\big|+\frac{r^2}{\log r}|\tr\overline{\chi}-\frac{2}{r}|
+r^{(\frac{5}{2}+\de)}|\hat{\overline{\chi}}|_{\ga}\right)+\nn\\
&&\sup_{C_0}\left[\left(\sum_{l=1}^q|r^{(2+l-\frac{2}{p})}\nabb^l\tr\chi|_{p,S}
+\sum_{l=1}^q|r^{(\frac{5}{2}+l+\de-\frac{2}{p})}\nabb^l\hat{\overline{\chi}}|_{p,S}
+\sum_{l=0}^q|r^{(\frac{7}{2}+l+\de-\frac{2}{p})}\nabb^l\dddd_4\hat{\overline{\chi}}|_{p,S}\right.\right.\nn\\
&&\left.\left.|r^{(2-\frac{2}{p})}\overline{\om}|_{p,S}+\sum_{l=1}^{q}|r^{(2+l+\de-\frac{2}{p})}\nabb^{l}\overline{\om}|_{p,S}\right)
+\sum_{l=1}^{q-1}|r^{(1+l+\de-\frac{2}{p})}\nabb^l\log\overline{\oom}|_{p,S}\right.\\
&&\left.+\sum_{l=0}^{q-1}|r^{(2+l-\frac{2}{p})}\nabb^l\overline{\ze}|_{p,S}+\sum_{l=0}^{q-2}|r^{(2+l-\frac{2}{p})}\nabb^{l}\overline{\omb}|_{p,S}
+\sum_{l=0}^{q-2}|r^{(1+l-\frac{2}{p})}\nabb^{l}\overline{\chib}|_{p,S}\right]\eql{Initialsmallcond1}\
.\nn
\eea 
\bea
&&J^{(q)}_{\Cb_0}\left[\overline{\ga}_{ab},\overline{\oomb},\overline{\underline X}^a,\overline{\ze}_a,\overline{\chi},\overline{\om}\right]=
\sup_{\Cb_0}\left(r\big|\overline{\oomb}-\frac{1}{2}\big|+\frac{r^2}{\log r}|\tr\overline{\chib}+\frac{2}{r}|
+r|\la|^{(\frac{3}{2}+\de)}|\hat{\overline{\chib}}|_{\ga}\right)+\eql{Initialsmallcond2}\nn\\
&&\sup_{\Cb_0}\left[\left(\sum_{l=1}^q|r^{(2+l-\frac{2}{p})}\nabb^l\tr\chib|_{p,S}
+\sum_{l=1}^q||\la|^{(\frac{3}{2}+\de)}r^{(1+l-\frac{2}{p})}\nabb^l\hat{\overline{\chib}}|_{p,S}
\right.\right.\nn\\
&&\left.\left.+\sum_{l=0}^q||\la|^{(\frac{5}{2}+\de)}r^{(1+l-\frac{2}{p})}\nabb^l\dddd_3\hat{\overline{\chib}}|_{p,S}
+|r^{(2-\frac{2}{p})}\overline{\omb}|_{p,S}+\sum_{l=1}^{q}|r^{(2+l+\de-\frac{2}{p})}\nabb^{l}\overline{\omb}|_{p,S}\right)
\right.\\
&&\left.+\sum_{l=1}^{q-1}|r^{(1+l+\de-\frac{2}{p})}\nabb^l\log\overline{\oomb}|_{p,S}
+\sum_{l=0}^{q-1}|r^{(2+l-\frac{2}{p})}\nabb^l\overline{\ze}|_{p,S}
+\sum_{l=0}^{q-2}||\la|^{\de}r^{((2-\de)+l-\frac{2}{p})}\nabb^{l}\overline{\om}|_{p,S}\right.\nn\\
&&\left.+\sum_{l=0}^{q-2}|r^{(1+l-\frac{2}{p})}\nabb^{l}\tr\overline{\chi}|_{p,S}
+\sum_{l=0}^{q-2}|r^{(2+l-\frac{2}{p})}\nabb^{l}\hat{\overline{\chi}}|_{p,S}
+\sum_{l=0}^{q-1}|r^{(1+l-\frac{2}{p})}\nabb^l\overline{\underline X}|_{p,S}\right]\ .\nn
\eea
where $p\in[2,4]$, $\de>0$ and $q\geq 7$.
\medskip

Observe that the first sum of norms is denoted by
$J^{(q)}_{C_0}\left[\overline{\ga}_{ab},\overline{\oom},\overline{\ze}_a,\overline{\chib},\overline{\omb}\right]$ where not all the
quantities appearing in the explicit expression are expressed. In fact we do not write those ones which are obtained on $C_0$ just
deriving along the outgoing direction, mainly $\chi$ and $\om$. The same holds for the second sum with the role of the various quantities
interchanged. 

\subsubsection{The spacetime solution norms}\label{SSSA.2} 
These norms are the same as those used in \cite{Kl-Ni:book}. We refer therefore, to sections 3.5.2 and 3.5.3, pages 90-95 of this
reference.
\bigskip

\NI {\bf Acknowledgments:} {\em One of the authors, (F.N.), want to thank P.Chr\'ushiel, H.Friedich and A.Ashtekar for some
interesting discussions. A particular thanks goes also to P.Chr\'ushiel, H.Friedich and P.Tod for the organization of the semester
on ``Global Problems in Mathematical Relativity", held at the Isaac Newton Institute for Mathematical Sciences, where part of this work was
completed.}
\newpage
\section{Figures}
\begin{figure}[hbt]
\centering 
\includegraphics[width=10.5cm]{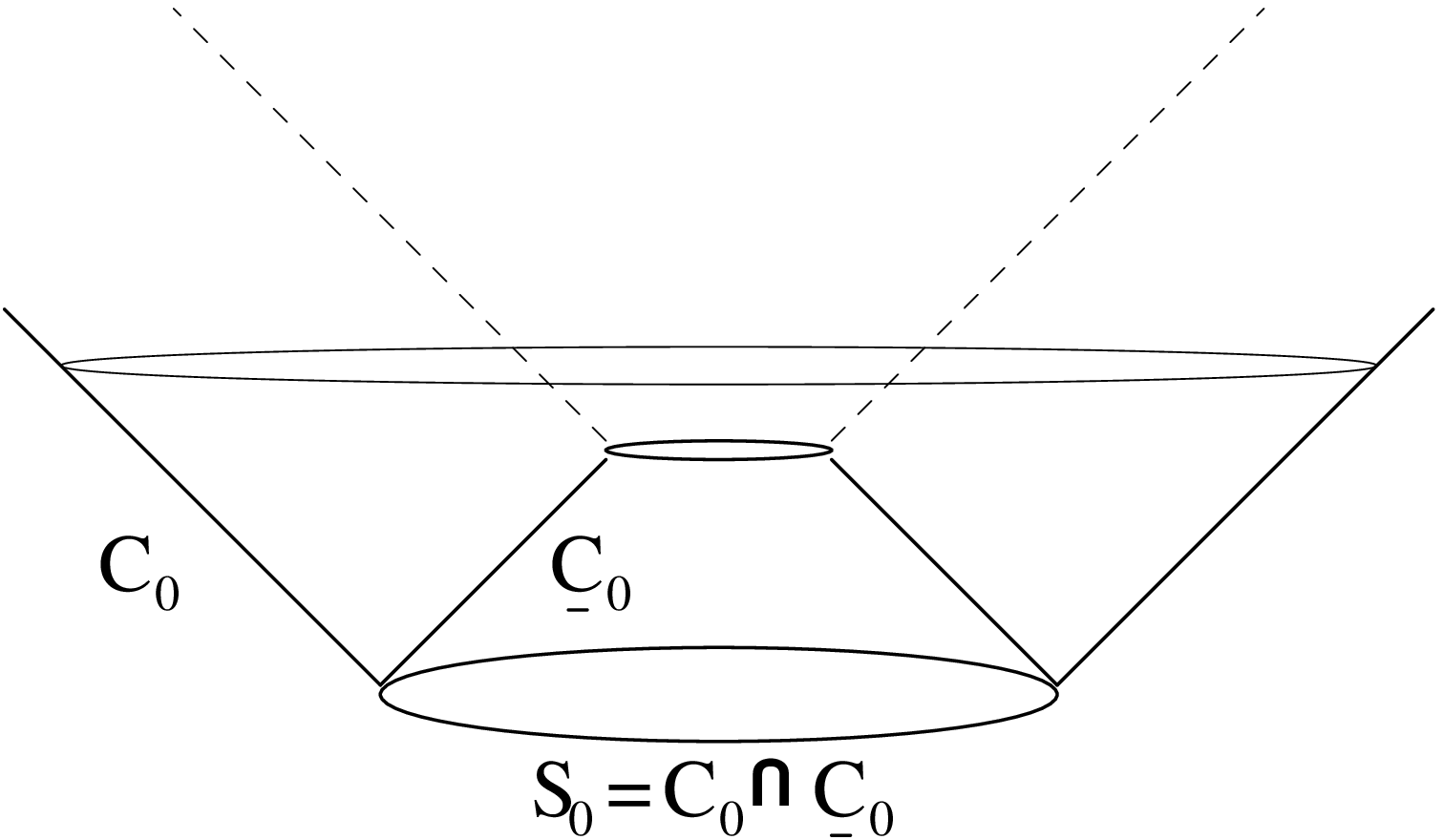}\label{fg1}
\caption{}
\end{figure}
\newpage
\begin{figure}[hbt]
\centering 
\includegraphics[width=10.5cm]{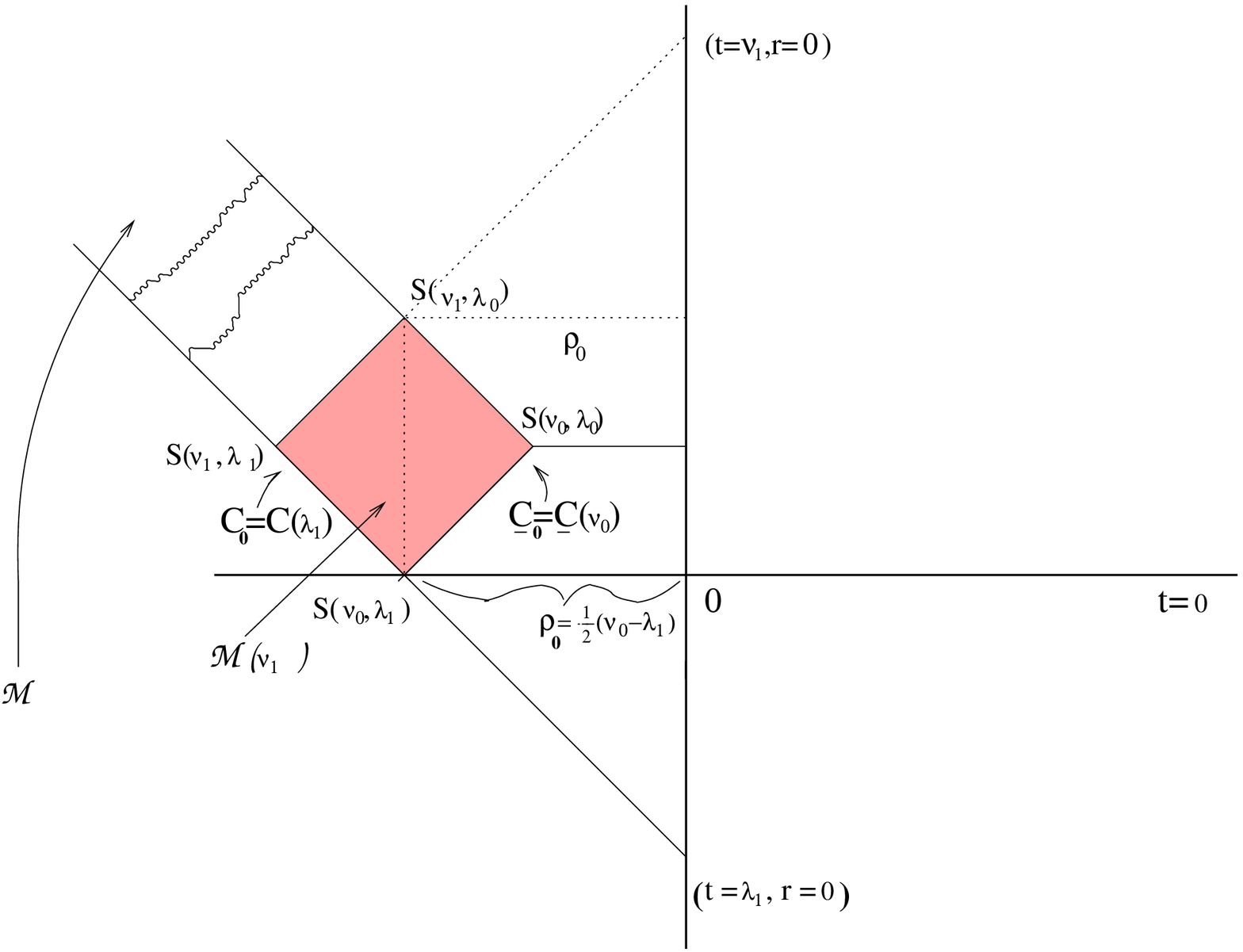}\label{fg2}
\caption{}
\end{figure}
\newpage
\begin{figure}[hbt]
\centering 
\includegraphics[width=10.5cm]{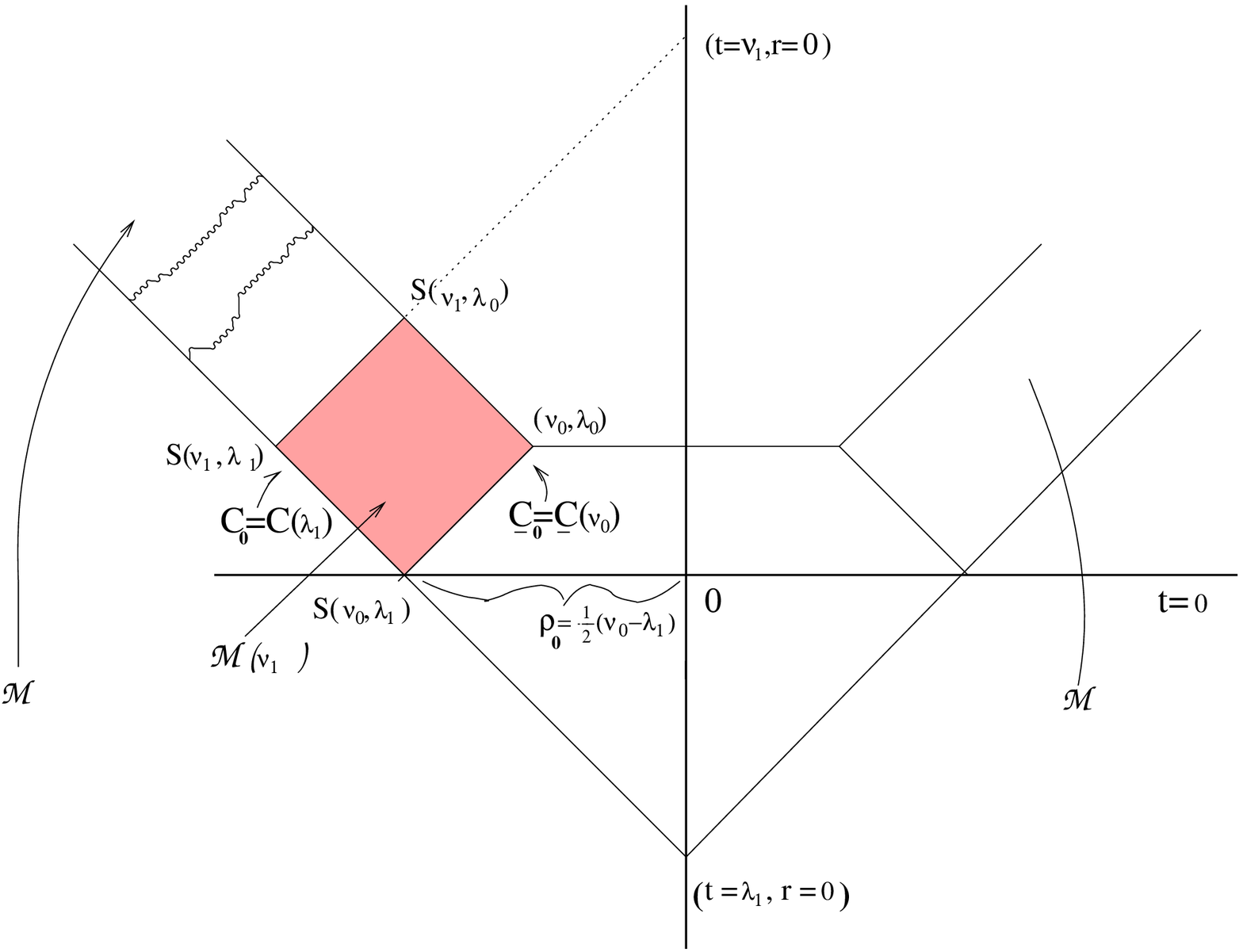}\label{fg3}
\caption{}
\end{figure}
\newpage
\begin{figure}[hbt]
\centering 
\includegraphics[width=7.5cm]{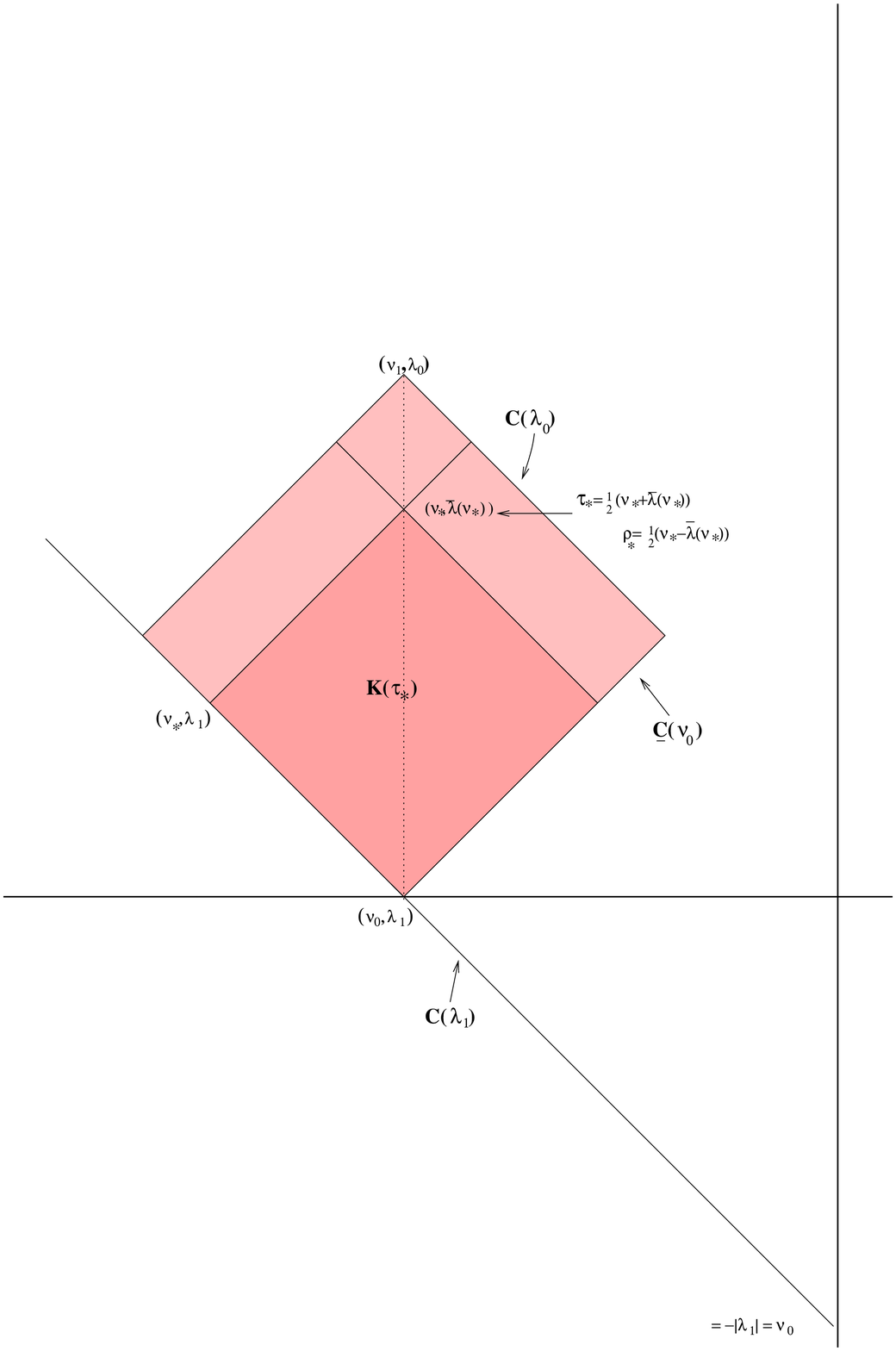}\label{fg4}
\caption{}
\end{figure}
\newpage
\begin{figure}[hbt]
\centering 
\includegraphics[width=10.5cm]{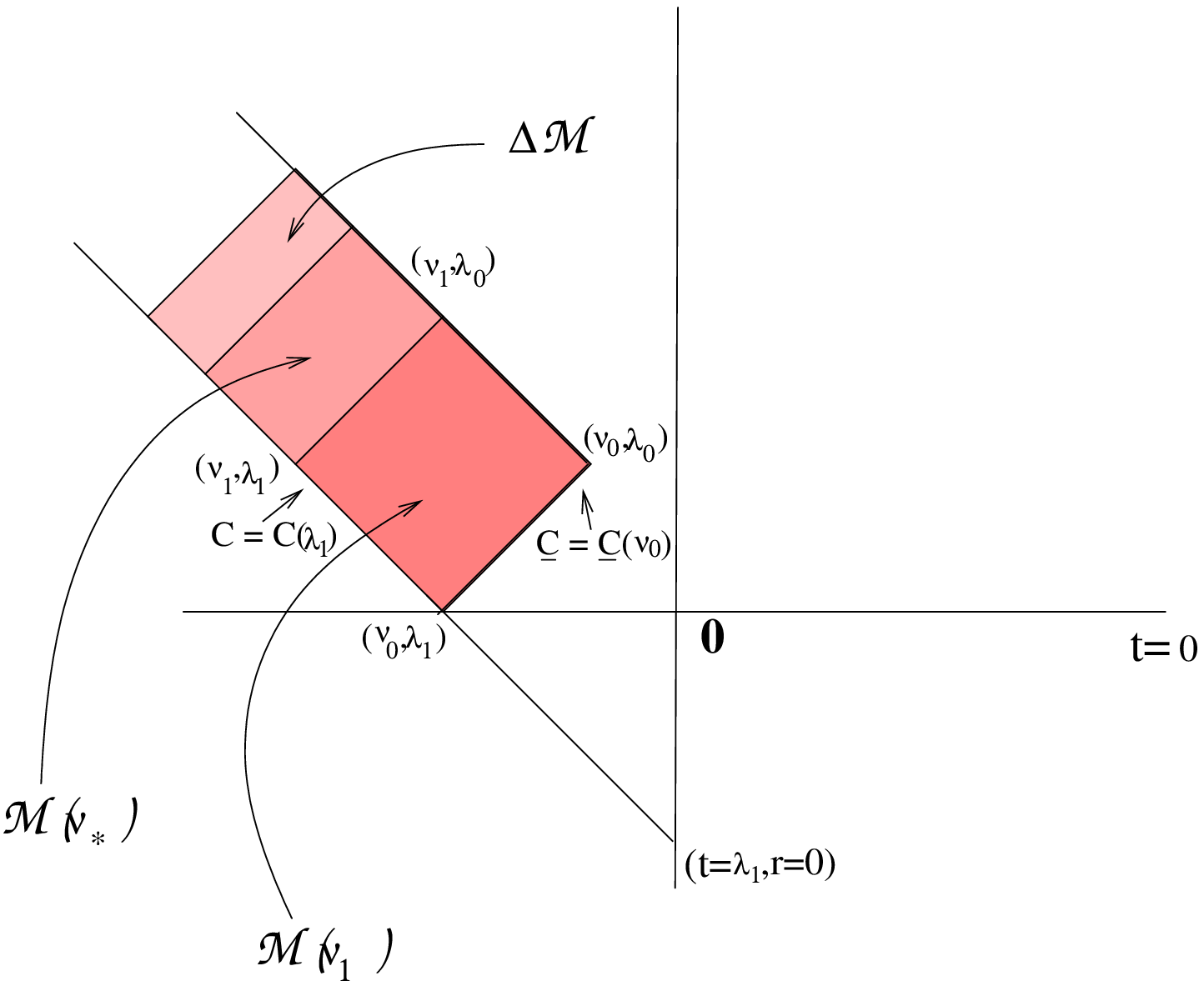}\label{fg5}
\caption{}
\end{figure}
\newpage
\begin{figure}[hbt]
\centering 
\includegraphics[width=7.5cm]{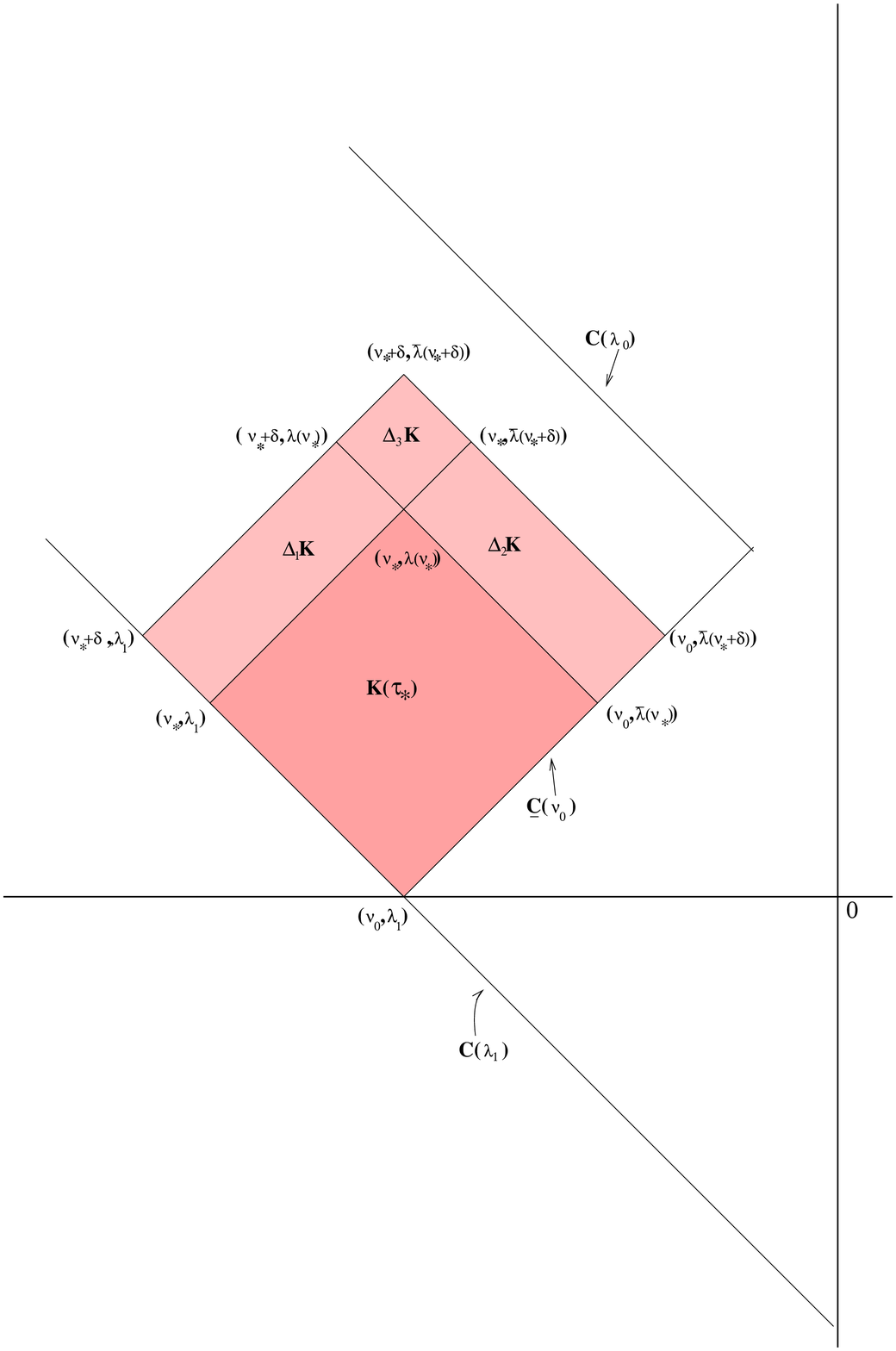}\label{fg6}
\caption{}
\end{figure}
\newpage
\begin{figure}[hbt]
\centering 
\includegraphics[width=7.5cm]{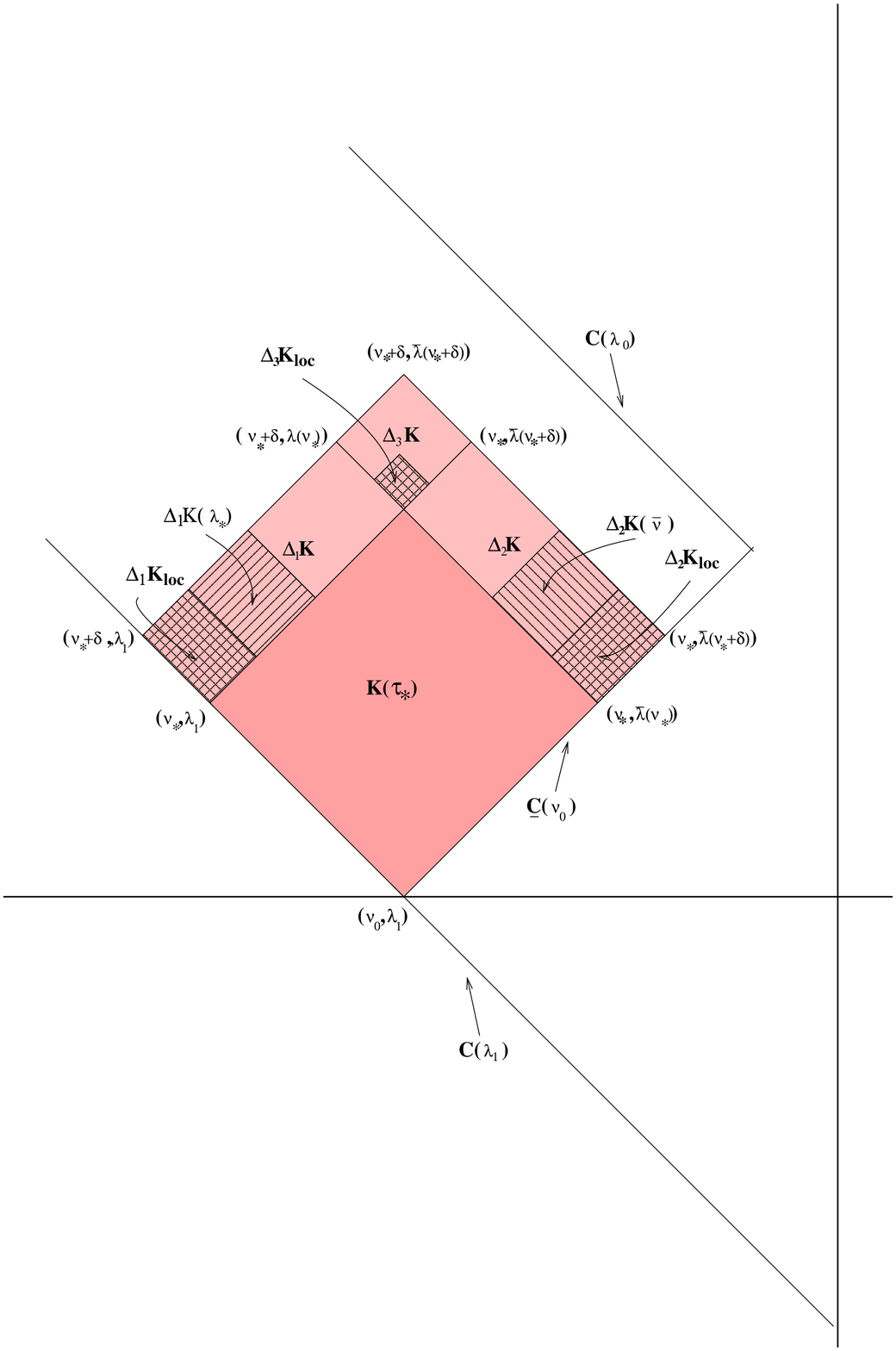}\label{fg7}
\caption{}
\end{figure}
\newpage

\bibliographystyle{math}
\bibliography{math}

\end{document}